 \documentclass[default,iicol,sn-mathphys-num]{sn-jnl}

\usepackage{graphicx}%
\usepackage{multirow}%
\usepackage{amsmath,amssymb,amsfonts}%
\usepackage{amsthm}%
\usepackage{longtable}
\usepackage{hyperref}
\usepackage{comment}
\usepackage{mathrsfs}%
\usepackage{pdflscape}
\usepackage{adjustbox}
\usepackage[title]{appendix}%
\usepackage{siunitx} 
\usepackage{xcolor}%
\usepackage{textcomp}%
\usepackage{manyfoot}%
\usepackage{braket}
\usepackage{chemfig}
\usepackage{tikz}
\usepackage{array}
\usepackage{quantikz}
\usepackage{booktabs}%
\usepackage{algorithm}%
\usepackage{algorithmicx}%
\usepackage{algpseudocode}%
\definecolor{blue_text}{RGB}{0,0,255}
\definecolor{red_text}{RGB}{255,0,0}
\usepackage{listings}%
\usepackage{float}

\newcommand{\LRG}{2\times10^{-4}}      
\newcommand{\LRD}{1\times10^{-4}}      
\newcommand{\BETAone}{0.5}
\newcommand{\BETAtwo}{0.9}
\newcommand{\BATCHSIZE}{64}
\newcommand{\NCRITIC}{5}               
\newcommand{\NITER}{2000}              
\newcommand{\LAMBDAgp}{10}
\newcommand{\PLSET}{\{1,2,4\}}         
\newcommand{\LAYERSET}{\{1,2,3,4\}}    

\theoremstyle{thmstyleone}%
\theoremstyle{thmstyletwo}%

\theoremstyle{thmstylethree}%

\begin{document}

\title[Hybrid QGANs for Drug Design]{Quantum-Classical Generative Models for Drug Design}


\author[1]{\fnm{Prateek} \sur{Jain}\textsuperscript{\textdagger}}\email{prateek.jain@fractal.ai}

\author[2]{\fnm{Param} \sur{Pathak}}

\author[2]{\fnm{Krishna} \sur{Bhatia}}

\author[1]{\fnm{Shalini} \sur{Devendrababu}}\email{shalini.devendrababu@fractal.ai}

\author[3]{\fnm{Srinjoy} \sur{Ganguly}}\email{srinjoyganguly@gmail.com}

\affil[1]{\orgdiv{\textit{QuantumAI Lab, Fractal Analytics}}, \orgaddress{\city{Gurugram}, \state{Haryana}, \postcode{122003}, \country{India}}}

\affil[2]{\orgdiv{\textit{QuantumAI Lab, Fractal Analytics}}, \orgaddress{\city{Mumbai}, \state{Maharashtra}, \postcode{400063}, \country{India}}}

\affil[3]{\orgdiv{\textit{Department of Physics and Astronomy, University College London}}, \orgaddress{\city{London}, \country{United Kingdom}}}


\abstract{
In molecular research, the modelling and analysis of molecules through simulation is an important part that has a direct influence on medical development, material science and drug discovery. The processing power required to design protein chains with hundreds of peptides is huge. Classical computing techniques, including state-of-the-art machine learning models being deployed on classical computing machines, have proven to be inefficient in this task, though they have been successful in a limited way. Moreover, current practical implementations, as opposed to purely theoretical modelling, are often infeasible in terms of both time and cost. One of the major areas where quantum machine learning is expected to have a profound advantage over classical algorithms is drug discovery. Quantum generative models have given some promising benefits in recent studies. This paper introduces three novel quantum generative adversarial network (QGAN) architecture variants resulting from different configurations, various quantum circuit layers and patched ansatz. A quantum simulator from Xanadu's PennyLane was utilized for executing the QGAN models trained on the QM9 dataset. Upon evaluation, one of the models, namely the QWGAN-HG-GP (Wasserstein distance with gradient penalty) model, outperformed the other QGAN models in different drug molecule property metrics.}




\maketitle

\renewcommand{\thefootnote}{\fnsymbol{footnote}}
\footnotetext{\textdagger Deceased}

\section{Introduction}\label{sec1}

Generative models are a kind of deep learning algorithms that are adapted to learn the underlying structures, patterns and distributions in the training data, thus allowing them to generate novel, comparable data points. Unlike traditional models that focus on predicting outputs from inputs, generative models seek to create entirely new samples that might resemble the training data. This is particularly useful in the complex process of drug discovery with the context of quantum chemistry and medical sciences. \textit{De novo} designing of drugs requires exploration of the chemical space which is astronomically large \cite{article}, far beyond the reach of brute‐force enumeration. Properties such as solubility, toxicity, binding affinity, lipophilicity; and the relationships between these properties and molecular features are quite efficiently captured by generative machine learning architectures \cite{article2}, \cite{article3}, \cite{article4}.

\subsection{Motivation}

Due to the exponentially large size of the chemical space, classical deep generative models often struggle with poor scalability, mode collapse, and an inability to capture quantum effects critical to molecular structure and reactivity.

Early research in quantum machine learning suggests that generative models supported by parametrized quantum circuits (PQCs) have the potential to represent complex probability distributions and navigate high-dimensional search spaces more efficiently than their classical counterparts. In this work, we attempt to bridge this gap by proposing architectures that combine quantum generators with classical discriminators to exploit the representational power of quantum models.

Furthermore, recent developments have further expanded the toolkit for novel molecular generation by integrating advanced generative architectures with quantum and classical computing paradigms. Auxiliary Discriminator Sequence Generative Adversarial Networks (ADSeqGAN) leverage an auxiliary discriminator to guide few‐sample molecule generation, improving diversity and fidelity in scarce data regimes \cite{tang2025adseqgan}. Bridging quantum and classical computing, transformer‐based GAN architectures have been proposed to harness the representational power of both domains for enhanced molecular design \cite{smith2025boqgan}. In parallel, comprehensive reviews of quantum machine learning applications in drug discovery have highlighted how hybrid quantum‐classical models can accelerate hit identification and optimization across academia and industry \cite{smaldone2025qml}. Most recently, quantum‑computing‑enhanced algorithms have demonstrated practical success by unveiling potential KRAS inhibitors, underscoring the real‐world impact of quantum generative methods in targeting challenging oncogenic proteins \cite{vakili2025quantum}.

\subsection{Key Contributions}

We connect quantum computing (QC) and generative drug design through hybrid quantum-classical adversarial learning frameworks, establishing the feasibility of quantum-enhanced molecule generation on near-term devices. Our key contributions are:

\begin{itemize}
    \item Three novel QGAN architectures—QWGAN-HG, QGAN-HG-GP, and QWGAN-HG-GP, that combine quantum generators with classical discriminators, incorporating Wasserstein distance and gradient penalties to stabilize adversarial training in molecular generation.
    
    \item Design a modular, patched PQC with variable-depth and parallelism to mitigate barren-plateau phenomena while maintaining expressibility in molecular graph generation.
    
    \item Demonstrate performance improvements of our models over classical MolGAN and baseline QGANs on the QM9 dataset across multiple evaluation metrics, including Fréchet distance, Wasserstein distance, and domain-specific drug-likeness properties (NP score, QED, LogP, SA score).
    
    \item Provide an end-to-end framework using PennyLane, RDKit, and PyTorch, enabling reproducible quantum-enhanced generative modeling pipelines for real-world drug discovery scenarios under NISQ-era (Noisy Intermediate Scale - era) constraints.
\end{itemize}
\section{Background and Related Work}
\label{bgr}

\subsection{Computational Drug Discovery and Data-Driven Generation}
Structure-based drug design relies on experimentally determined 3D macromolecular structures curated in resources such as the Protein Data Bank \cite{article25}. While physics-based molecular simulation provides mechanistic insight into atomic motions and energetics \cite{article26}, recent work increasingly complements these approaches with \emph{data-driven generators} trained on curated molecular corpora. In this study we follow the latter paradigm: we learn to model the distribution of small organic molecules using \textbf{QM9} for training/evaluation and compare against drug-like reference distributions from \textbf{ZINC}/\textbf{ZINC-250k} where appropriate \cite{qm9,zinc15}. 

\begin{figure}[t]
  \centering
  \includegraphics[scale=.5]{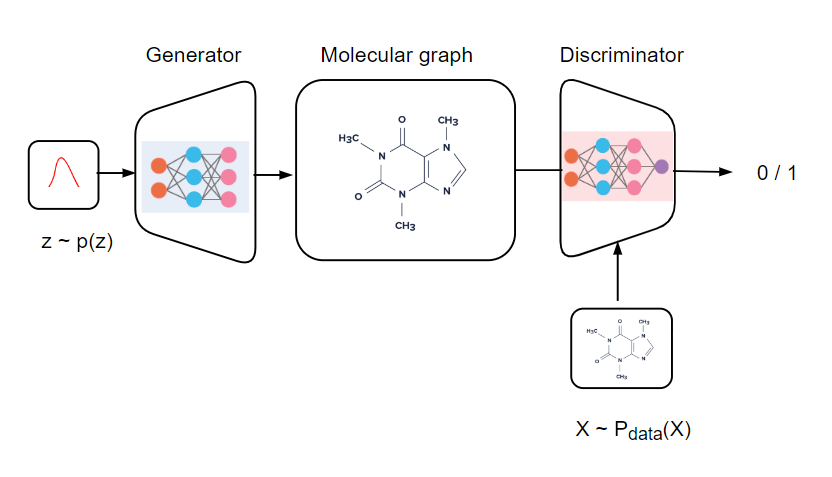}
  \caption{\textbf{Graph GAN for molecule generation.} A generator maps latent vectors to molecular graphs (node features for atoms; edge tensors for bond types). A discriminator contrasts generated graphs with samples from a reference dataset, driving the generator to match the real distribution and produce novel, valid, drug-like structures.}
  \label{fig:molgan}
\end{figure}

\subsection{GANs for Molecular Generation}
\paragraph{Implicit graph GANs.}
\textit{MolGAN} introduced implicit adversarial training directly on molecular graphs, with a multilayer perceptron (MLP) generator outputting an $N{\times}N{\times}B$ bond tensor and an $N{\times}F$ atom-feature matrix, and a GCN-based discriminator assessing realism \cite{molgan}. An auxiliary reward head provides valency/connectivity signals via a REINFORCE-style update, avoiding expensive rejection sampling. On QM9, MolGAN reported $>$80\% validity and $>$70\% uniqueness while highlighting mode-collapse risks without stabilization.

\paragraph{Wasserstein and gradient-penalty variants.}
To improve stability, many graph GANs adopt the Wasserstein objective with a gradient penalty (WGAN-GP). With critic $D$ and generator $G$, the standard loss is
\begin{align}
\mathcal{L}_D
  &= \mathbb{E}_{x\sim P_r}[D(x)]
   - \mathbb{E}_{z\sim P_z}[D(G(z))] \nonumber \\
  &\quad + \lambda\,\mathbb{E}_{\hat{x}\sim P_{\hat{x}}}
     \!\left[\left(\lVert\nabla_{\hat{x}} D(\hat{x})\rVert_2 - 1\right)^2\right].
\end{align}

which enforces an approximate 1-Lipschitz critic and mitigates gradient pathologies. MedGAN exemplifies this line by pairing WGAN(-GP) training with GCN-based graph encoders/decoders for scaffold-focused generation \cite{macedo2024medgan}.

\paragraph{Sequence (SMILES) GANs with policy gradients.}
For discrete SMILES, gradients cannot backpropagate through sampling. \textit{SMILES-MaskGAN} addresses this with an actor--critic policy-gradient update
\begin{equation}
\begin{split}
\nabla_\theta J(\theta)
= \mathbb{E}_{y_t\sim G_\theta}\!\Bigg[
\sum_{t=1}^T\!\Big(\sum_{i=t}^{T}\gamma^{i} r_i - b_t\Big)\,\\
\qquad \nabla_\theta \log G_\theta(y_t)
\Bigg].
\end{split}
\end{equation}

where rewards $r_i$ derive from a discriminator and $b_t$ is a variance-reduction baseline \cite{lee2021generative}. Masked token prediction stabilizes training and improves uniqueness/novelty on GuacaMol distribution benchmarks.

\paragraph{Target-focused adversarial design.}
Beyond unconditional distribution learning, adversarial objectives have also been applied to \emph{target-specific} design. For example, \cite{bian2019deep} train dcGANs over molecular fingerprints with CNN discriminators (LeNet/AlexNet/VGG) to bias generation toward cannabinoid receptor activity, achieving strong AUCs in retrospective classification. Although fingerprint-based, these works illustrate how adversarial learning interfaces with virtual-screening pipelines.
\subsection{Classical Generative Baselines on QM9 and ZINC-250k}
\label{sec:classical-baselines}

\paragraph{SMILES language models.}
Transformer decoders learn next-token distributions over SMILES and scale well to large corpora (e.g., \textit{MolGPT}) \cite{molgpt}. Fragment-aware tokenization (e.g., \textit{t\text{-}SMILES}) improves syntactic robustness and distribution fidelity on QM9/ZINC-style chemistry \cite{tsmiles}.

\paragraph{Graph variational autoencoders (VAEs).}
Graph VAEs map molecular graphs to continuous latents and decode atoms/bonds under valence constraints. \textit{JT-VAE} factorizes generation into a scaffold (junction) tree and attachments, enabling high validity and competitive optimization on QM9 and ZINC-250k \cite{jtvae}.

\paragraph{Normalizing-flow generators.}
Invertible flows enable exact likelihoods and efficient sampling. \textit{GraphNVP} introduced coupling flows for graphs on QM9/ZINC \cite{graphnvp}. \textit{GraphAF} combined autoregressive factorization with flows and strong valence handling, reporting high validity and property alignment on ZINC-250k \cite{graphaf}. \textit{MoFlow} extended invertible bijections to molecular graphs with competitive ZINC-250k performance \cite{moflow}. Residual- and discrete-flow variants such as \textit{GRF} and \textit{GraphDF} improve expressivity or handle discrete latents directly, with comprehensive evaluations on the same benchmarks \cite{grf,graphdf}.

\paragraph{Adversarial (GAN) models on graphs.}
\textit{MolGAN} adapts GAN training to molecular graphs (with optional RL fine-tuning) and demonstrated high validity on QM9, while noting mode-collapse risks without stabilization \cite{molgan}. Although less prevalent today than flows/transformers, graph GANs remain a common reference in ZINC/QM9 comparisons.

\paragraph{Masked/denoising graph modeling.}
Masked Graph Modeling (MGM) learns conditional distributions by masking nodes/edges and iteratively predicting the masked parts; it achieves competitive FCD/KL and property alignment on QM9 and larger chemistry (e.g., ChEMBL), with straightforward transfer to ZINC-style evaluations \cite{mgm}.

\paragraph{Evaluation practice.}
Across QM9 and ZINC-250k, standard practice reports (i) \emph{validity}, \emph{uniqueness}, and \emph{novelty}; (ii) property distributions (e.g., QED, logP, MW, SA); and (iii) distributional metrics such as Fréchet ChemNet Distance (FCD) and KL divergence \cite{fcd,guacamol}. Our comparisons follow these conventions and cite widely adopted configurations from the original papers \cite{molgpt,tsmiles,jtvae,graphnvp,graphaf,moflow,grf,graphdf,molgan}.

\begin{table}[t]
\centering
\small
\caption{Representative classical generative baselines on QM9 / ZINC-like datasets.}
\label{tab:classical_sota}
\begin{tabular}{@{}llp{0.58\columnwidth}@{}}
\toprule
Model & Data & Representative results \\
\midrule
JT-VAE~\cite{jtvae} 
  & ZINC-250k 
  & Reports essentially perfect validity and uniqueness and good alignment of QED, logP and SA distributions. \\
GraphAF~\cite{graphaf} 
  & ZINC-250k 
  & High validity and uniqueness with strong KL scores on QED, logP and SA. \\
MoFlow~\cite{moflow} 
  & ZINC-250k 
  & Competitive validity and uniqueness with favourable property-matching on QED, logP and SA. \\
MGM~\cite{mgm} 
  & QM9 
  & High validity and uniqueness with competitive KL/FCD on graph statistics. \\
MolGPT~\cite{molgpt} 
  & MOSES (ZINC) 
  & Very high validity and uniqueness with Fr\'echet ChemNet Distance around $0.07$. \\
\bottomrule
\end{tabular}
\end{table}

\subsection{QGAN with a Hybrid Generator}

In classical drug discovery, computational approaches face overwhelming bottlenecks when navigating chemical spaces estimated to contain $10^{60}$ to $10^{100}$ chemically feasible molecules. Traditional GANs suffer from training instabilities, mode collapse, and inability to explore certain regions of chemical space, particularly when tasked with generating small molecules requiring extensive parameterization. Li \textit{et al.} \cite{li2021quantum}, addressed these limitations by developing QGAN-HG (Quantum GAN with Hybrid Generator), a novel architecture that leverages quantum computing's intrinsic probabilistic nature to overcome classical GAN constraints. The fundamental challenge arose from the exponential qubit requirements of full quantum GANs, where generating QM9-like molecules would demand over 90 qubits ($\log_2 5^{36} + \log_2 5^9 > 90$ qubits). The proposed hybrid approach employs a quantum generator coupled with a classical discriminator, optimized through the standard minimax objective:

\begin{equation}
\begin{split}
\min_{\theta_g} \max_{\theta_d} V(D, G) 
&= \mathbb{E}_{x \sim p_{\text{data}}(x)}[\log D(x)] \\
&\quad + \mathbb{E}_{z \sim p_z(z)}[\log(1 - D(G(z)))],
\end{split}
\end{equation}

where $\theta_g$ and $\theta_d$ represent the generator and discriminator parameters respectively, $D(x)$ is the discriminator's probability assessment that input $x$ is real, $G(z)$ is the generator output from noise input $z$, $p_{\text{data}}(x)$ is the real data distribution, and $p_z(z)$ is the noise distribution. A QGAN-HG architecture, originally proposed in the work being discussed, contains a generator that is a parametrized quantum circuit (PQC) accompanied by a classical layer of neural networks as the discriminator. The PQC, provides a feature vector. The classical layer is then utilized to provide a vector representing discrete atoms and a matrix denoting bonds between these atoms for depicting the drug molecules in the form of a graph.

The QGAN-HG demonstrated remarkable parameter efficiency, achieving comparable molecular generation quality with only 1.97\% of the original classical GAN parameters while maintaining similar Fréchet Distance (FD) scores and drug property metrics. The hybrid architecture successfully generated valid drug molecules with appropriate QED (quantitative estimate of druglikeness), logP (solubility), and SA (synthetic accessibility) scores. This quantum-enhanced approach not only circumvented the exponential scaling requirements of full quantum implementations but also provided inherent protection against gradient vanishing problems through shortened neural network depths, establishing a new paradigm for quantum-classical hybrid generative models in computational drug discovery.

Upon contemplating classical GANS, one can witness fundamental training instabilities arise from mode collapse, generator-discriminator oscillations, and convergence difficulties stemming from the mathematical properties of minimizing Jensen-Shannon divergence. QGANs inherited these classical adversarial learning challenges while introducing additional quantum-specific limitations including excessive quantum parameter requirements and circuit depth constraints in the NISQ era. Existing quantum approaches like PQWGAN suffered from resource inefficiency, requiring 28 subgenerators with substantial quantum gate overhead, while maintaining the inherent instability of the adversarial minimax formulation: $\min_G \max_D E_{x\sim p_{data}(x)}[\log D(x)] + E_{z\sim p_z(z)}[\log(1-D(G(z)))]$, where quantum generators struggled with amplitude-based pixel value generation requiring zero amplitudes for completely dark pixels.

Ma \textit{et al.} \cite{ma2025quantum}, addressed these fundamental limitations through the Quantum Implicit Neural Representation-based Quantum Generative Adversarial Network (QINR-QGAN) architecture that leverages quantum data re-uploading circuits combined with Wasserstein distance optimization. The QINR framework exploits the mathematical equivalence between data re-uploading quantum circuits and Fourier Neural Networks, where each quantum circuit component can be expressed as:
\begin{align}
f(h) &= \langle 0|U^\dagger(h)OU(h)|0\rangle, \\
U(h) &= W^{(L)}S(h)W^{(L-1)}...W^{(2)}S(h)W^{(1)}, \\
f(h) &= \sum_{K,J} \alpha_{K,J}e^{i(K-J)\cdot h},
\end{align}
where $S(h) = e^{-ih_1H}\otimes...\otimes e^{-ih_{d_h}H}$ represents the encoding layers and the final expression demonstrates the Fourier series equivalence with frequency spectrum $\{K-J\}_{K,J}$. The QINR-QGAN training employs WGAN-GP stabilization through the objective:

\begin{align}
\min_G \max_{D \in \mathcal{D}} \; &\mathbb{E}_{x \sim P_{\text{data}}} [D(x)] 
- \mathbb{E}_{z \sim P_z(z)} [D(G(z))] \notag \\
&- \lambda \, \mathbb{E}_{\hat{x} \sim P_{\hat{x}}} 
\left[ \left( \| \nabla_{\hat{x}} D(\hat{x}) \|_2 - 1 \right)^2 \right],
\end{align}

where the gradient penalty term enforces the 1-Lipschitz constraint and $P_{\hat{x}}$ represents uniform sampling between generator and data distributions.

Experimental validation across MNIST, Fashion-MNIST, and EMNIST datasets demonstrated that QINR-QGAN achieved performance comparable to state-of-the-art models while utilizing nearly 10 times fewer trainable quantum parameters (72 vs 1008-3600 parameters compared to PQWGAN and Quantum AnoGAN). The architecture successfully generated high-resolution images up to 78×64 pixels on the CelebA dataset, with evaluation metrics showing FID scores competitive with Quantum AnoGAN, superior SSIM performance across all baseline models, and stable training convergence with Wasserstein distances approaching zero. The quantum resource efficiency demonstrates particular promise for NISQ-era applications, maintaining image generation quality while dramatically reducing quantum computational overhead and circuit complexity.

\subsection{Quantum Wasserstein GAN (QWGAN)}
In classical molecular design, generative adversarial networks face significant computational bottlenecks, particularly in generating diverse and chemically valid molecular structures. Classical MolGAN architectures suffer from training instability and limited expressivity, often converging to a "high entropy state" where the generator produces disconnected atoms or simple molecular fragments rather than complex, drug-like compounds. The traditional drug design process requires approximately 15 years and \$1 billion per compound, with classical GANs achieving limited diversity in the generated chemical space. Anoshin \textit{et al.} \cite{anoshin2024hybrid}, addressed these limitations by introducing hybrid quantum cycle generative adversarial networks that integrate PQCs into classical molecular generation frameworks. The approach employs variational quantum circuits (VQCs) as the initial layer of the generator, utilizing either vanilla variational repetitive quantum (VVRQ) or exponential fourier quantum (EFQ) configurations. The generator operates through a composition of single-qubit rotation gates defined as $\text{Rot}(\theta_1, \theta_2, \theta_3) = R_y(\theta_1) \cdot R_z(\theta_2) \cdot R_y(\theta_3)$, where each component rotation is specified by $R_y(\theta) = \begin{pmatrix} \cos(\theta/2) & -\sin(\theta/2) \\ \sin(\theta/2) & \cos(\theta/2) \end{pmatrix}$ and $R_z(\theta) = \begin{pmatrix} e^{-i\theta/2} & 0 \\ 0 & e^{i\theta/2} \end{pmatrix}$; where quantum entanglement is facilitated through CNOT gates applied between adjacent qubits. The hybrid framework combines Wasserstein GAN loss with cycle consistency and reward components through the unified objective: $L(\theta) = \lambda \cdot L(\theta)_{\text{WGAN}} + (1-\lambda) \cdot L(\theta)_{\text{Cycle}} + \gamma L(\theta)_{\text{Reward}}$, where $\lambda$ and $\gamma$ regulate the relative contributions of each component. This quantum-classical hybrid approach demonstrated substantial improvements, achieving up to 30\% increases in quantitative estimation of druglikeness (QED) scores and enhanced synthesis accessibility metrics compared to classical baselines. The integration of cycle components effectively mitigated the high entropy state problem, enabling generation of more chemically diverse and realistic molecular structures while maintaining training stability across quantum machine learning frameworks.

Classical conditional molecular generative models face significant scalability challenges due to high-dimensional input constraints and training instability when optimizing with standard GAN objectives. In particular, conditional GANs (cGANs) for SMILES string generation often suffer from mode collapse and sparse reward landscapes, making conditional property control unreliable and sample quality inconsistent across conditions.

To address this, Kan \textit{et al.}~\cite{pajuhanfard2024survey}, proposed \emph{QwQGAN}, a hybrid quantum-classical generative model that replaces the classical generator with a PQC and employs the Wasserstein GAN loss. The PQC receives as input both quantum-encoded noise and classical molecular conditions. The generator circuit is composed of a data reuploading structure followed by entanglement layers, and the training uses a dual-objective quasi-Wasserstein loss to stabilize updates. The hybrid critic $D$ is trained via a smoothed Wasserstein loss:

\begin{equation}
\begin{split}
\mathcal{L}_{\text{critic}} 
&= \mathbb{E}_{x \sim P_r}[D(x)] 
- \mathbb{E}_{z \sim P_z}[D(G(z \mid c))] \\
&\quad + \lambda \, \mathbb{E}_{\hat{x}} \left[ 
\left( \|\nabla_{\hat{x}} D(\hat{x})\|_2 - 1 \right)^2 
\right],
\end{split}
\end{equation}

where $c$ is the conditioning input, $G(z \mid c)$ is the quantum generator output conditioned on $c$, and $\hat{x}$ is a linear interpolation between real and generated samples. The use of parameterized rotations and entanglement gates allows expressive generation while keeping the quantum depth shallow enough for NISQ-era implementation.
QwQGAN was evaluated on the MOSES benchmark, where it achieved superior performance in terms of novelty (0.926), validity (0.984), and internal diversity compared to baseline cGAN and cVAE models. The integration of quantum encoding with conditional control helped generate molecules that not only matched target properties but also maintained structural variety. The model also exhibited training stability benefits due to the quasi-Wasserstein loss formulation, mitigating vanishing gradients and improving convergence. This work demonstrates that quantum-enhanced conditional generators can be practically deployed for fine-grained molecular design tasks, especially under low-data and constrained-representation regimes.

In classical GANs, fundamental convergence issues arise from mode collapse, oscillatory behavior, and non-unique Nash equilibria, particularly when the discriminator performs suboptimal measurements that lead to indefinite oscillation between generator states. Previous quantum GAN architectures (QuGAN) directly analogized classical GAN frameworks but suffered from similar instabilities, where the optimal Helstrom measurement operator becomes orthogonal to true quantum data and opposite to generated data, causing the generator to overshoot the target state and subsequently oscillate. The minimax optimization in QuGAN: $\min_{\theta_g} \max_T (\text{Tr}[T\sigma] - \text{Tr}[T\rho(\theta_g)])$ yields linear discriminator functions that prove non-optimal for quantum state certification problems. Niu \textit{et al.} \cite{niu2022entangling}, addressed these fundamental limitations through the Entangling Quantum GAN (EQ-GAN) architecture that leverages quantum entanglement between both generator output and true quantum data rather than evaluating them individually. The EQ-GAN discriminator employs a parameterized destructive ancilla-free swap test that enables fidelity-based measurements between quantum states, with the core discriminator operation defined as:

\begin{equation}
D_\sigma^{\text{fid}}(\rho(\theta_g)) = \text{Tr}\left[\sigma^{1/2} \rho(\theta_g) \sigma^{1/2}\right]^2,
\end{equation}

\begin{align}
D_\sigma(\theta_d, \rho(\theta_g)) 
&= \frac{1}{2} \Big[ 1 
+ \cos^2\theta_d \notag \\
&\quad + \sin^2\theta_d \cdot D_\sigma^{\text{fid}}(\rho(\theta_g)) \Big],
\end{align}

\begin{equation}
\min_{\theta_g} \max_{\theta_d} V(\theta_g, \theta_d) = \min_{\theta_g} \max_{\theta_d} \left[1 - D_\sigma(\theta_d, \rho(\theta_g))\right],
\end{equation}

where the swap test circuit ansatz $U(\theta_d) = \exp[-i\theta_d \text{CSWAP}]$ enables matrix exponentiation of controlled swap gates. The architecture guarantees convergence to the global optimal Nash equilibrium when $\rho(\theta_g) = \sigma$, with the discriminator circuit transforming input states through: $HU(\theta_d)H|0\rangle_a|\psi\rangle|\zeta\rangle = \frac{i\sin\theta_d}{2}|1\rangle_a[|\zeta\rangle|\psi\rangle - |\psi\rangle|\zeta\rangle] + \frac{1}{2}|0\rangle_a[(e^{-i\theta_d} + \cos\theta_d)|\psi\rangle|\zeta\rangle - i\sin\theta_d|\zeta\rangle|\psi\rangle]$. Experimental validation on Google Sycamore quantum processors demonstrated superior convergence properties with state fidelity errors of $(0.6 \pm 0.2) \times 10^{-4}$ compared to $(2.4 \pm 0.5) \times 10^{-4}$ for perfect swap tests, while achieving 69\% accuracy on variational quantum random access memory (QRAM) tasks and successful applications in quantum neural network training where classical approaches failed to converge.

\subsection{Alternative Quantum Approaches and Positioning}\label{sec:alt-q}

Beyond QGANs, multiple quantum paradigms have been explored for molecular design. Here we briefly position our hybrid QGANs relative to two prominent alternatives.

\paragraph{Quantum annealing / QUBO formulations.}
Discrete molecular tasks (fragment linking, constrained scaffold assembly, docking pose selection) can be cast as quadratic unconstrained binary optimization (QUBO) problems and solved via quantum annealing or hybrid quantum–classical tabu/SA solvers. These approaches are appealing when objectives (e.g., fragment compatibility, steric clashes, simple property surrogates) are naturally combinatorial and low order. Reported case studies generally operate on small graphs or coarse fragment sets and often require strong problem reductions; scalability and cost remain active constraints on current hardware. In contrast, our QGANs learn \emph{distributions} over molecular graphs and thus natively support sampling for diversity, while still admitting hard constraints through the critic and post-hoc RDKit validation. 

\paragraph{Quantum circuit Born machines (QCBM) and related generative priors.}
QCBMs directly model a probability distribution via a parameterized quantum state whose amplitudes define sample likelihoods. They offer elegant likelihood-free training but typically require careful circuit/measurement design to avoid barren plateaus and to scale beyond small molecule vocabularies. Recent hybrids use QCBMs as priors or feature maps within classical decoders. Compared to QCBMs, our adversarial training with Wasserstein loss and gradient penalty empirically improves stability, and patched PQCs trade depth for parallel expressivity that is friendlier to NISQ constraints. 

\paragraph{Takeaway.}
Alternative quantum approaches are attractive for tightly constrained, small-scale combinatorial subproblems or as priors; our hybrid QGANs target \emph{distribution learning at scale} with stability features (Wasserstein + GP) and architectural choices (patched PQCs) that reduce depth while preserving expressivity. We view the methods as complementary: annealing/QUBO for discrete micro-decisions, QCBM for priors, and hybrid QGANs for high-throughput sample generation under drug-likeness criteria.

\section{Theoretical Background \& Mathematical Preliminaries}

\subsection{GANs and Wasserstein Distance}

Generative Adversarial Networks (GANs) constitute a powerful generative modeling framework introduced by Goodfellow et al., where a generator \( G \) learns to map samples from a latent space \( z \sim P_z \) to the data space, while a discriminator \( D \) attempts to distinguish real samples \( x \sim P_r \) from generated ones \( G(z) \sim P_g \). The two models are trained in a minimax game, given by:

\begin{equation}
\min_G \max_D \mathbb{E}_{x \sim P_r}[\log D(x)] + \mathbb{E}_{z \sim P_z}[\log(1 - D(G(z)))].
\end{equation}

where \( x \) is a real data point sampled from the true distribution \( P_r \), \( z \) is a latent variable sampled from a prior distribution \( P_z \), \( G(z) \) is the generator’s output, and \( D(\cdot) \) denotes the discriminator's scalar prediction.

However, training instability and vanishing gradients have motivated alternative formulations, notably the Wasserstein GAN (WGAN). WGAN replaces the Jensen-Shannon divergence with the Earth Mover (EM) or Wasserstein-1 distance, defined between probability distributions \( P_r \) and \( P_g \) over a metric space \( \mathcal{X} \) as:

\begin{equation}
W(P_r, P_g) := \inf_{\gamma \in \Pi(P_r, P_g)} \mathbb{E}_{(x,y) \sim \gamma} [\|x - y\|],
\end{equation}

where \( \gamma \) is a joint distribution over \( x \) and \( y \) such that its marginals are \( P_r \) and \( P_g \), and \( \|x - y\| \) denotes the Euclidean distance between a real sample \( x \) and a generated sample \( y \).

Due to computational intractability of directly optimizing this primal form, the Kantorovich-Rubinstein duality reformulates it as:

\begin{equation}
W(P_r, P_g) = \sup_{\|f\|_L \leq 1} \mathbb{E}_{x \sim P_r}[f(x)] - \mathbb{E}_{x \sim P_g}[f(x)],
\end{equation}

where \( f \) is any 1-Lipschitz function, representing the critic (or discriminator) constrained to have gradients bounded by 1.

In practice, the discriminator \( D \) is constrained to satisfy the Lipschitz condition, enforced via gradient penalty or weight clipping. The WGAN objective thus becomes:

\begin{equation}
\mathcal{L}_{\text{WGAN}} = \mathbb{E}_{x \sim P_r}[D(x)] - \mathbb{E}_{z \sim P_z}[D(G(z))],
\end{equation}

where \( D(x) \) is the critic score for real data \( x \), and \( D(G(z)) \) is the score for fake data produced by the generator from latent input \( z \).

with an additional penalty term introduced in WGAN-GP (Gradient Penalty), defined as:

\begin{equation}
\mathcal{L}_{\text{GP}} = \lambda \, \mathbb{E}_{\hat{x} \sim P_{\hat{x}}} \left[ (\|\nabla_{\hat{x}} D(\hat{x})\|_2 - 1)^2 \right],
\end{equation}

where \( \hat{x} \) is an interpolated sample between real and generated data, \( \nabla_{\hat{x}} D(\hat{x}) \) is the gradient of the critic with respect to \( \hat{x} \), and \( \lambda \) is a hyperparameter controlling the strength of the penalty.

This penalty term enforces the 1-Lipschitz constraint $|\nabla D(\mathbf{x})|_2 \approx 1$ throughout the input space by explicitly penalizing deviations from unit gradient norm. Unlike weight clipping (used in the original WGAN), gradient penalty prevents unbounded discriminator gradients that cause training instability. It also avoids capacity limitations from hard constraints on weights; provides smoother, more informative gradients to the generator. This approach also directly helps prevent mode collapse because it keeps the gradients flowing smoothly, even when the model's output and the real data don't overlap. This is especially important for our hybrid quantum-classical systems, where the quantum part of the generator can get stuck in vast, flat areas that offer no useful signals. By ensuring gradients remain stable, we can actually train the model effectively. In essence, this work provides the core theory needed to use Wasserstein-based loss functions in QGANs. It creates a more reliable training process and ensures the model can generate a wide variety of outputs, which is absolutely vital for creating molecules in complex fields like drug discovery.

\subsection{Quantum Generator Circuit and Training Protocol}

Our quantum generator is implemented as a patched variational circuit that maps a classical latent vector \( \mathbf{z}\in\mathbb{R}^d \) into an $N$‑qubit quantum state, whose measured outcomes are post‑processed by a shallow classical network.
Our patched PQC, as shown in Fig.~\ref{pqc}, uses small fixed-size blocks of $Q$ qubits (e.g. $Q=2$), tiled in parallel into $P\in\PLSET$ patches. Each patch applies an encoding block followed by $L\in\LAYERSET$ variational layers of single-qubit rotations and nearest-neighbour entanglers. The best-performing QWGAN-HG-GP configuration uses $P=4$ patches and $L=2$ layers (P4–L2), balancing expressivity and circuit depth.

\subsubsection{State Encoding}
Each latent dimension \(z_i\) is angle‑encoded onto one qubit via the composite rotation:

\begin{equation}
U_{\rm init}(z_i)
=\;R_Y\bigl(\pi\,z_i\bigr)\;
R_Z\bigl(2\pi\,z_i\bigr)\;
R_Y\bigl(\pi\,z_i\bigr)\,.
\end{equation}

This prepares the qubit in a superposition whose amplitude and phase both depend on \(z_i\).  This choice balances hardware‑friendliness (only single‑qubit rotations) with expressivity in mapping \(\mathbb{R}\to\mathrm{SU}(2)\).

\subsubsection{Ansatz Structure.} We stack $L$ variational layers, where each layer consists of:
\begin{itemize}
  \item \textbf{Parameterized rotations.}  A bank of single‑qubit \(R_Y(\theta_{j,\ell})\) gates acting on each qubit \(j=1\ldots N\) in layer \(\ell\).
  \item \textbf{Entangling block.}  For each neighboring pair \((j,j+1)\), a CNOT gate, then a phase rotation \(R_Z(\phi_{j,\ell})\) on qubit \(j+1\), then a second CNOT gate, forming a CNOT–\(R_Z\)–CNOT ``sandwich.''
  \item \textbf{Phase shifts.}  A secondary layer of single‑qubit $R_Z(\phi_{j,\ell})$ gates to enrich the relative phases introduced by entanglement.
\end{itemize}
Altogether, each layer introduces $2N$ real trainable parameters, so the total trainable parameters per $L$‑layer patch is \(V_{\rm patch}=2NL\).  For a 4‑qubit, 3‑layer circuit this yields \(V_{\rm patch}=24\).  With $M$ patches in parallel (to encode $d=M$ latent features) the full generator uses $V=M\,V_{\rm patch}$ parameters.

\subsubsection{Circuit Depth and Noise Considerations.} The depth of each patch is
\[
D_{\rm patch} \;=\; 3 \;+\; L\,(1_{\!R_Y}+1_{\!{\rm CZ}}+1_{\!R_Z}) \;\approx\; 3 + 3\cdot3 = 12
\]
layers of primitive gates.  In our simulations we found $L=3$ to hit the best trade‑off between expressivity (no barren‑plateau onset) and resilience to depolarizing noise at $p\leq 0.01$.

\paragraph{Measurement and Read‑out.} After the final layer we apply single‑qubit $Z$‑basis measurements.  The bitstring outcomes \(\mathbf{m}\in\{0,1\}^N\) are aggregated into a real vector via expectation values $\langle Z_j\rangle = \Pr(m_j=0)-\Pr(m_j=1)$.  These $N$ expectation values enter a two‑layer classical perceptron that outputs the discriminator’s input features.

\subsubsection{Training Details.} We train the QGAN in the standard adversarial fashion:
\begin{itemize}
  \item \textbf{Optimizer:} Adam with $\beta_1=0.5,\ \beta_2=0.9$.
  \item \textbf{Learning rate:} $\alpha_G = 2\times10^{-4}$ for the generator, $\alpha_D = 1\times10^{-4}$ for the discriminator.
  \item \textbf{Batch size:} 64 latent samples per update.
  \item \textbf{Epochs:} 2{,}000 iterations over the QM9 training split.
  \item \textbf{Loss functions:}  Wasserstein loss with gradient penalty (weight $\lambda=10$) as in \cite{article21}.
\end{itemize}

\subsection{Molecular Representation}

We adopt a graph-based encoding of molecules as atom- and bond-feature matrices, in order to bridge quantum outputs and chemical structures. Atoms are indexed by integer types (e.g., carbon = 1, nitrogen = 2, oxygen = 3), forming a feature vector \( \mathbf{B} \), while bonds (single = 1, double = 2) populate an adjacency matrix \( \mathbf{A} \in \mathbb{R}^{N \times N} \). 

Each entry \( A_{ij} \) encodes the bond order between atoms \( i \) and \( j \), and \( B_i \) specifies the atom identity. This discrete graph is ultimately reconstructed from the quantum-generated feature vectors via Gumbel-Softmax discretization, enabling seamless integration with classical discriminators and property-evaluation tools such as \texttt{RDKit}.

\section{Methodology \& Experimental Setup}

\subsection{Simulator and Hardware Overview}

All quantum circuits were executed on Xanadu’s PennyLane CPU-based simulator (default \texttt{"default.qubit"} device), interfaced via the PennyLane-PyTorch plugin. Simulations ran on a 16-core CPU environment, with each PQC evaluated in batched mode (batch size = 32) to parallelize circuit executions. 

For potential hardware deployment, the same PennyLane scripts can target IBM Quantum backends through the \texttt{pennylane-qiskit} plugin, with noise models and transpilation routines defined per backend. 

\subsection{Proposed Framework}
In this part, we draw parallels between the existing generative models - the classical GAN as well as the quantum GAN models and introduce variations in the hybrid quantum-classical GAN structure. We put forward three new architectures namely Quantum-Wasserstein GAN along with a Hybrid Generator (QWGAN-HG), Quantum GAN with Hybrid Generator and Gradient Penalty (QGAN-HG-GP) and Quantum-Wasserstein GAN with Hybrid Generator and Gradient Penalty (QWGAN with HG and GP). We analyse the operation and output of these proposed model architectures on the Pennylane CPU-based Quantum Simulators \cite{article22}. Furthermore, we explain the intricate technicalities of the proposed models: 

\subsubsection{QWGAN-HG} 

\begin{figure*}
  \centering
  \includegraphics[width=1\linewidth]{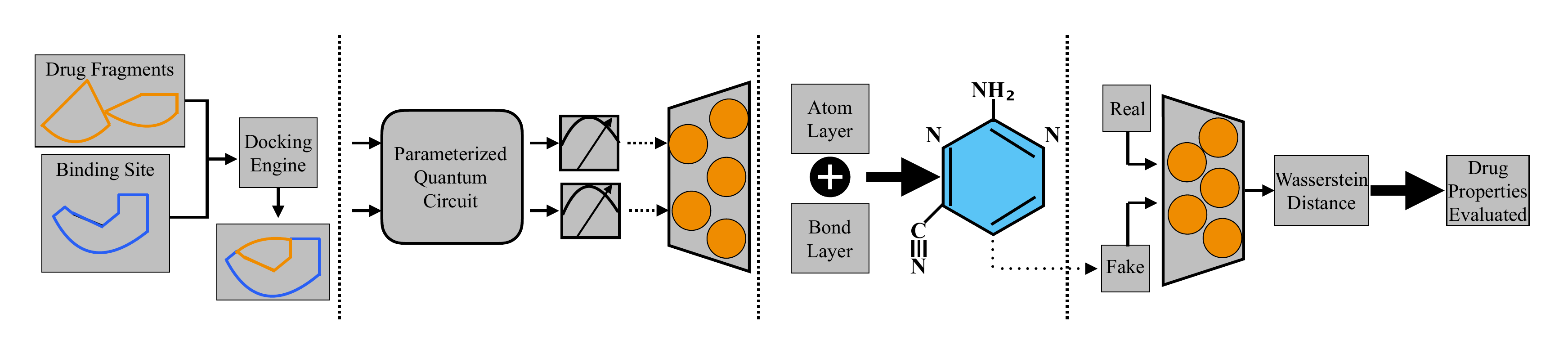}
  \vspace{-0.8cm}
  \caption{Quantum Wasserstein-GAN Hybrid Generator (QWGAN-HG), incorporates a Wasserstein distance metric to enhance molecular generation stability. The PQC processes binding site data and transforms it into feature vectors, which classical atom and bond layers convert into molecular graphs for evaluation against real QM9 dataset structures.}
  \label{QWGANHG}
 \end{figure*}

As shown in Fig.~\ref{QWGANHG}, the QWGAN-HG architecture represents our first proposed enhancement over the base QGAN-HG model. This architecture incorporates a hybrid structure where a PQC processes inputs from drug fragments and binding sites through the docking engine. The quantum circuit utilizes quantum gates to transform these inputs into a feature vector. This feature vector is then forwarded as an input into two classical components—the atom layer and bond layer—which work together to construct a complete molecular structure. The molecule representation is then compared against real molecules from the QM9 dataset, with the discriminator (illustrated as a cloud with orange circles) computing the Wasserstein distance to evaluate the generator's performance.

The PQC $U(\boldsymbol{\theta})$ is defined as a sequence of quantum operations:

\begin{equation}
U(\boldsymbol{\theta}) = \prod_{l=1}^{L} \left( \bigotimes_{i=1}^{N} R_Y(\theta_{l,i}) \cdot \text{Entangler} \right),
\end{equation}

where the number of quantum circuit layers is represented by $L$, the number of qubits is represented by $N$, $R_Y(\theta_{l,i})$ is the single qubit rotation gate applied to qubit $i$ in layer $l$, and Entangler Multi-qubit entangling gates CNOT, for instance, to induce correlations.

The generator $G(z; \boldsymbol{\theta})$ maps latent noise $z \sim p(z)$ to molecular graphs through the PQC, producing atom and bond matrices. The critic $D$, implemented as a classical neural network, computes the Wasserstein distance $W$ between the generated distribution $\mathbb{P}_g$ and real data distribution $\mathbb{P}_r$:

\begin{equation}
W(\mathbb{P}_r, \mathbb{P}_g) = \sup_{\|f\|_L \leq 1} \mathbb{E}_{x \sim \mathbb{P}_r}[f(x)] - \mathbb{E}_{z \sim p(z)}[f(G(z))],
\end{equation}

where $\|f\|_L \leq 1$ enforces a 1-Lipschitz constraint on the critic function $f$. Unlike traditional GAN approaches that may suffer from training instabilities, the QWGAN-HG utilizes the continuity of the Wasserstein metric to provide smoother gradients during training. This mitigates common issues such as mode collapse and vanishing gradients in quantum neural networks (QNNs), enabling stable convergence and improved modeling of molecular distributions. The framework excels at capturing complex drug-like features, including binding affinity and chemical stability, by aligning the synthetic data distribution with the QM9 dataset through adversarial optimization.

\subsubsection{QGAN-HG-GP}

\begin{figure*}
  \centering
  \includegraphics[width=1\linewidth]{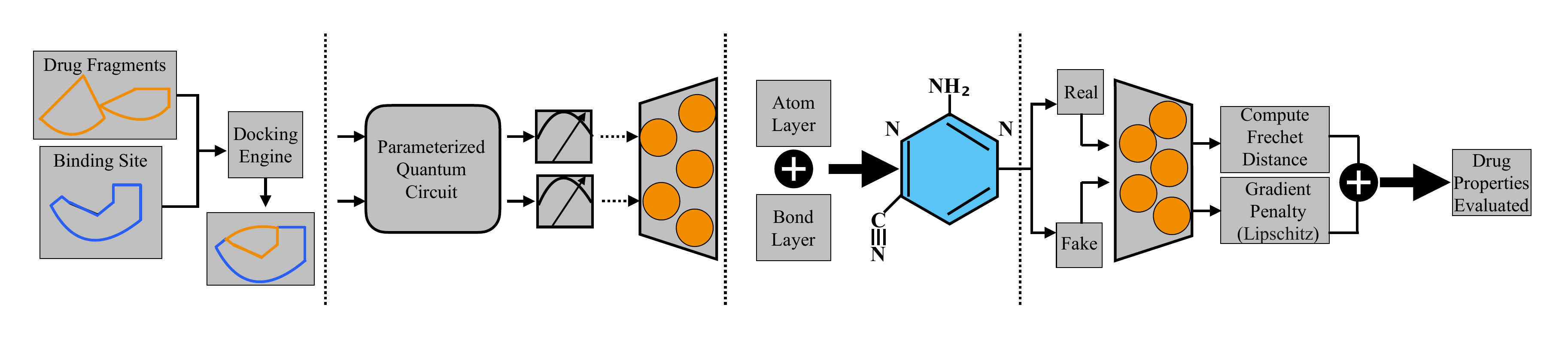}
  \vspace{-0.8cm}
  \caption{Quantum GAN Hybrid Generator with Gradient Penalty (QGAN-HG-GP), an enhanced hybrid architecture that augments the base QGAN-HG model with gradient-penalty regularization for improved training stability.}
  \label{QGANHGGP}
 \end{figure*}

The QGAN-HG-GP architecture Fig.~\ref{QGANHGGP} enhances the base QGAN-HG model by comprising a gradient penalty (GP) into the critic’s loss function, stabilizing training through Lipschitz continuity enforcement. Drug fragments and binding site data are first processed by a docking engine to generate molecular candidates, which are encoded into quantum feature vectors via a PQC. These vectors are decoded by classical atom and bond layers to construct molecular graphs.

The critic evaluates both real molecules (from QM9) and generated molecules by computing the Wasserstein distance, augmented with a gradient penalty term:

\begin{align}
\mathcal{L}_{\text{critic}} = \ 
&\underbrace{
\mathbb{E}_{\tilde{x} \sim \mathbb{P}_g}[D(\tilde{x})] 
- 
\mathbb{E}_{x \sim \mathbb{P}_r}[D(x)]
}_{\text{Wasserstein term}} \notag \\
&+ \lambda \cdot \mathbb{E}_{\hat{x} \sim \mathbb{P}_{\hat{x}}}
\left[
\left( \left\lVert \nabla D(\hat{x}) \right\rVert_2 - 1 \right)^2
\right],
\end{align}

where $\mathbb{P}_r$ and $\mathbb{P}_g$ are the real and generated distributions, $\mathbb{P}_{\hat{x}}$ interpolates linearly between them, and $\lambda$ (typically set to 10) balances the penalty term. By penalizing deviations from the 1-Lipschitz constraint ($\|\nabla D\|_2 \approx 1$), the critic avoids unstable weight clipping and mitigates vanishing gradients, enabling smoother convergence.

Unlike classical GANs, the hybrid quantum-classical generator makes use of PQC’s ability to model high-dimensional molecular features efficiently, while the critic’s GP term ensures robust adversarial training. Validated molecules are then assessed for drug-like properties (NP, QED, LogP), with the architecture outperforming baseline models in both distribution alignment (lower Fréchet/Wasserstein distances) and pharmaceutical viability.

\subsubsection{QWGAN-HG-GP} 

\begin{figure*}
  \centering
  \includegraphics[width=1\linewidth]{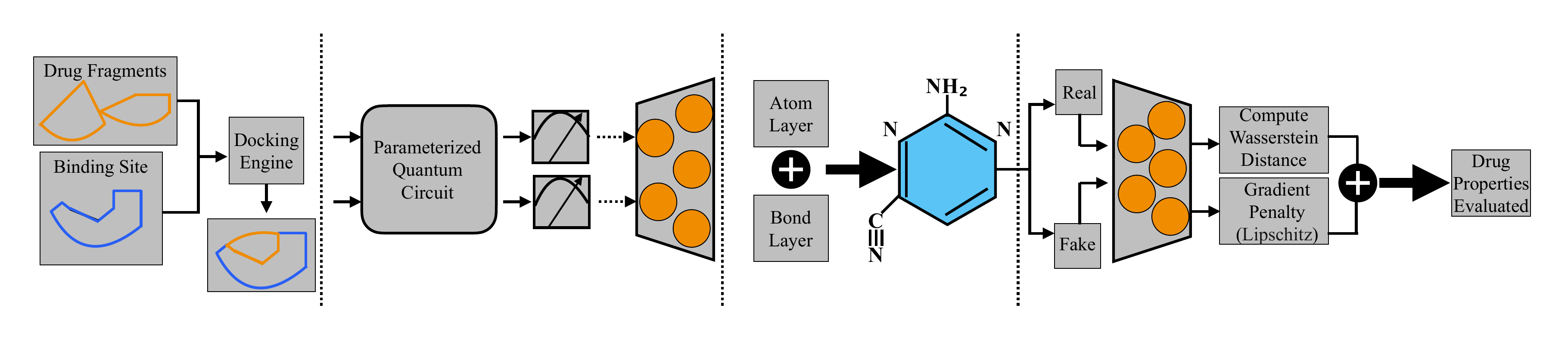}
  \vspace{-0.8cm}
  \caption{Quantum Wasserstein-GAN Hybrid Generator with Gradient Penalty (QWGAN-HG-GP), the most advanced hybrid architecture combining Wasserstein distance metrics with gradient penalty regularization for optimal molecular generation. This model implements a specialized L4, P2 quantum circuit configuration that divides computation into parallel entangled qubit groups, effectively mitigating barren plateaus while enabling efficient exploration of high-dimensional chemical spaces with minimal qubit requirements.}
  \label{QWGANHGGP}
 \end{figure*}

The QWGAN-HG-GP architecture, as shown in Fig.~\ref{QWGANHGGP}, refines the hybrid quantum-classical framework by combining Wasserstein distance, gradient penalty (GP), and a 4-layered (L4), 2-patched (P2) quantum circuit to enhance molecular generation and training stability. Drug fragments and binding site data are preprocessed by a docking engine, producing molecular candidates that are encoded into quantum feature vectors via a PQC). The PQC is structured into 2 parallel patches (P2), each containing 4 variational layers (L4) of entangled qubits, enabling efficient exploration of high-dimensional chemical spaces while mitigating barren plateaus.
These quantum feature vectors are decoded by classical atom and bond layers to construct molecular graphs (e.g., nodes NH$_2$, bonds). The critic enforces Lipschitz continuity through gradient penalties on interpolated samples $\hat{x} \sim \mathbb{P}_{\hat{x}}$, ensuring stable adversarial training. Validated molecules undergo rigorous drug property evaluation, with QWGAN-HG-GP (L4–P2) outperforming all baselines in synthesizing chemically valid, drug-like candidates.

\subsection{Implementation Details}

The implementation consists of a hybrid quantum-classical framework built on PennyLane for quantum circuit simulation, PyTorch for classical neural networks, and RDKit for molecular validation and property calculation. Key libraries include \textit{frechetdist} for Fréchet distance computation and pennylane-qiskit for optional IBM Quantum backend integration.

The adversarial training process employs the Adam optimizer for both generator and discriminator, chosen for its adaptive learning rates and robustness to noisy gradients, with
$\beta_1=\BETAone$ and $\beta_2=\BETAtwo$, to balance momentum and gradient variance. Other hyperparamters used in our models, are listed in table~\ref{tab:hyperparams}. Unless otherwise noted, we use generator learning rate $\LRG$,
critic learning rate $\LRD$, batch size \BATCHSIZE, \NCRITIC\ critic updates per generator step,
and a total of \NITER\ training iterations per run.
The PQC as shown in Fig.~\ref{pqc} initializes qubits with $R_Y$ and $R_Z$ rotations, followed by repeated layers of tunable single-qubit rotations $R_Y(\theta_i)$, $R_Z(\theta_j)$ and multi-qubit entangling gates for instance CNOT, to enhance expressibility. The parameters $\{\theta_i\}$ are optimized via classical methods to minimize task-specific costs, with final qubit measurements mapped to feature vectors for downstream tasks. For stability, the critic incorporates a gradient penalty, and the gradient penalty coefficient is fixed to
$\lambda_{\mathrm{gp}}=\LAMBDAgp$ throughout, to enforce a 1-Lipschitz constraint, replacing weight clipping and mitigating barren plateaus in quantum parameter optimization.
Quantum circuits are structured into \textit{patches} (p1, p2, p4), where p4 divides the circuit into four parallel sub-circuits with 2 qubits each, enhancing expressibility while reducing qubit requirements. Post-processing employs the Gumbel-Softmax trick to discretize molecular graphs, critical for handling categorical bond/atom representations. Each configuration was trained with 5 independent random seeds; we report mean metric values across these runs (variances are below 0.01 and available in the code)

\begin{figure*}[t]
    \centering
    \begin{quantikz}
    \lstick{$|0\rangle$} & \gate{RY(Z_1)} & \gate{RZ(Z_2)} & \gate{RY(\theta_1)} & \ctrl{1} & \qw & \qw & \meter{} \\
    \lstick{$|0\rangle$} & \gate{RY(Z_1)} & \gate{RZ(Z_2)} & \gate{RY(\theta_2)} & \targ{} & \gate{Z(\theta_{N+1})} & \ctrl{1} & \meter{} \\
    \lstick{$|0\rangle$} & \gate{RY(Z_1)} & \gate{RZ(Z_2)} & \gate{RY(\theta_3)} & \qw & \gate{Z(\theta_{N+2})} & \targ{} & \meter{} \\
    \lstick{$|0\rangle$} & \gate{RY(Z_1)} & \gate{RZ(Z_2)} & \gate{RY(\theta_4)} & \qw & \gate{Z(\theta_{N+3})} & \qw & \meter{}
    \end{quantikz}
    \caption{Schematic representation of the PQC structure used in our QGAN architectures, showing the initialization layer (green), variational layers (blue), and output feature vector mapping. The circuit's modular design enables flexible configuration into different patch arrangements (p1, p2, p4) and layer depths that significantly impact molecular generation performance.}
    \label{pqc}
\end{figure*}

\begin{table}[h!]
\centering
\caption{Canonical training hyperparameters for QGAN variants. Values that differ for ZINC are indicated explicitly.}
\begin{tabular}{@{}lcc@{}}
\toprule
\textbf{Setting}              & \textbf{QM9}              & \textbf{ZINC-20} \\
\midrule
Batch size                    & \BATCHSIZE               & \BATCHSIZE \\
Generator lr                  & $\LRG$                   & $\LRG$ \\
Critic lr                     & $\LRD$                   & $\LRD$ \\
Adam $(\beta_1,\beta_2)$      & $(\BETAone,\BETAtwo)$    & $(\BETAone,\BETAtwo)$ \\
Critic steps per G step       & \NCRITIC                 & \NCRITIC \\
Total iterations              & \NITER                   & \NITER \\
Gradient penalty $\lambda_{\rm gp}$ & $\LAMBDAgp$       & $\LAMBDAgp$ \\
Patches $P$                   & $\PLSET$                 & $\PLSET$ \\
Layers $L$                    & $\LAYERSET$              & $\LAYERSET$ \\
\bottomrule
\end{tabular}
\label{tab:hyperparams}
\end{table}

\begin{figure}[htbp]
\centering

\begin{minipage}[t]{0.45\textwidth}
\centering
\chemfig{*6((-N(-CH_3)-CH_3)=O-N=CH-C(=O)-N(-CH_3)-)}

\vspace{0.5cm}
\textbf{Molecular structure as graph}
\end{minipage}
\hfill
\begin{minipage}[t]{0.45\textwidth}
\centering

\[
A =
\begin{bmatrix}
0 & 1 & 0 & 0 & 0 & 0 & 0 & 0 & 1 \\
1 & 0 & 1 & 0 & 0 & 0 & 0 & 0 & 0 \\
0 & 1 & 0 & 0 & 0 & 0 & 1 & 0 & 0 \\
0 & 0 & 2 & 0 & 0 & 0 & 0 & 0 & 0 \\
0 & 0 & 0 & 0 & 0 & 0 & 0 & 0 & 0 \\
0 & 0 & 0 & 0 & 1 & 0 & 0 & 0 & 0 \\
0 & 0 & 0 & 0 & 0 & 1 & 0 & 1 & 0 \\
0 & 0 & 1 & 1 & 0 & 0 & 1 & 0 & 2 \\
1 & 0 & 0 & 0 & 0 & 0 & 0 & 2 & 0
\end{bmatrix}
\]

\vspace{0.4cm}

\[
B = \begin{bmatrix} 1 & 3 & 1 & 1 & 1 & 1 \end{bmatrix}
\]

\vspace{0.4cm}
\textbf{Adjacency Matrix (A) and Feature Vector (B)}
\end{minipage}

\caption{Representation of a caffeine-like molecule from the QM9 dataset illustrating the graph-based encoding used throughout our generative models. The adjacency matrix A captures bond connections (with values indicating bond types) while vector B encodes atom types, together forming the complete structural representation processed by our quantum-classical architectures.}
\label{fig:molmatrix}
\end{figure}

\subsection{The Datasets} 
\label{sec:dataset}

\subsubsection{The QM9 Dataset}
The \href{https://www.kaggle.com/datasets/zaharch/quantum-machine-9-aka-qm9}{QM9} dataset, is a fundamental resource for researchers in computational chemistry.  It documents quantum chemical properties for over 134,000 small, stable organic molecules that up to nine atoms, including carbon, hydrogen, oxygen, and nitrogen along with their 3D structural details. Beyond validation, QM9 aids in pioneering new methodologies like integrating ML with QC and streamlines research into how molecular structures correlate with their properties.
In a graph-based structure, molecules are shown as Atom and Bond vector matrices. Taking an example of a molecule in a graph based structure in Fig.\ref{fig:molmatrix}, a corresponding matrix based representation where N and 0 atom types are depicted by 2 and 3 in the vector counterpart of the atom. Single \& double bonds are depicted as 1 and 2, respectively, in the matrix counterpart of the bond.

\subsubsection{ZINC-20 (Drug-like scaling experiment)}
To probe scalability beyond small QM9 molecules (up to 9 heavy atoms), we construct a ``ZINC-20'' subset from the ZINC-250k drug-like collection~\cite{zinc15}. We parse SMILES strings with RDKit, add explicit hydrogens and retain only molecules with at most 20 heavy (non-hydrogen) atoms. Each molecule is featurized as a graph with a fixed node budget $N=20$, atom types drawn from the ZINC vocabulary in Sec.~\ref{sec:dataset}, and three bond channels (single, double, aromatic) as in our QM9 setup. This defines a controlled scaling task from QM9-scale molecules to larger, more drug-like structures while keeping the graph size fixed.

\subsection{Generated Distribution Evaluation Metrics}

To evaluate the effectiveness of generative models in capturing the intrinsic properties of training data distributions, the following metrics are employed:
\subsubsection{Fréchet Distance}
The statistical similarity between quantum-generated molecular distributions and authentic data is evaluated using the Fréchet distance \cite{article24}. This metric combines the disparity between mean vectors (\(\boldsymbol{\mu}\), \(\boldsymbol{\mu}_w\)) and covariance matrices (\(\boldsymbol{C}\), \(\boldsymbol{C}_w\)) as follows:

\begin{align}
d^2\left((\boldsymbol{\mu}, \boldsymbol{C}), (\boldsymbol{\mu}_w, \boldsymbol{C}_w)\right) 
&= \left\|\boldsymbol{\mu} - \boldsymbol{\mu}_w\right\|_2^2 \nonumber \\
&\quad 
+
\operatorname{Tr}\left(\boldsymbol{C} + \boldsymbol{C}_w - 2\left(\boldsymbol{C} \boldsymbol{C}_w\right)^{1/2}\right)
\end{align}

where \(\boldsymbol{\mu}_w\) and \(\boldsymbol{C}_w\) correspond to the mean and covariance of the QM9 dataset's molecular distribution. For computation, the QGAN synthesizes molecular batches (treated as high-dimensional points), while an identical number of reference molecules are randomly selected from QM9. The metric is computed across these batches, with lower values indicating greater similarity between generated and real molecular distributions.


\subsubsection{Wasserstein Distance}A feature of the Wasserstein \cite{article19} distance metric that allows GANs to correlate with picture quality produced by the generator is its presence. By minimizing generator loss, Wasserstein GAN's critic NN loss function encourages convergence as opposed to balance between the discriminator and generator. The Barren Plateaus, or gradient vanishing problem, is what the Wasserstein distance attempts to solve. To compute the Wasserstein-1 distance between two distributions $\mu$ and $\nu$, we define:
\begin{equation}
    \mathcal{W}_1(\mu, \nu) = \sup_{g \in \mathcal{F}} \left( \mathbb{E}_{\mathbf{u} \sim \mu} [g(\mathbf{u})] - \mathbb{E}_{\mathbf{v} \sim \nu} [g(\mathbf{v})] \right),
\end{equation}

where $\mathcal{F}$ is the set of all 1-Lipschitz continuous functions $g$, such that 
$|g(\mathbf{u}) - g(\mathbf{v})| \leq \|\mathbf{u} - \mathbf{v}\| \quad \text{for all } \mathbf{u}, \mathbf{v}$.

\subsubsection{Gradient Penalty} 

\( g \), a differentiable function, satisfies the 1 - Lipschitz continuity condition only if its gradient norm is universally bounded by 1, i.e., \( \|\nabla g(\boldsymbol{x})\|_2 \leq 1 \ \forall \boldsymbol{x} \). To enforce this constraint during adversarial training, a gradient penalty term is incorporated into the loss function, penalizing deviations from the target gradient norm. For a critic function \( g \), the composite loss is defined as:

\begin{align}
\mathcal{L}_{\text{gp}} =\ 
&\underbrace{\mathbb{E}_{\boldsymbol{u} \sim \mu}\left[g(\boldsymbol{u})\right] 
- \mathbb{E}_{\boldsymbol{v} \sim \nu}\left[g(\boldsymbol{v})\right]}_{\text{Wasserstein Term}} \nonumber \\
&+ \ \gamma \cdot \mathbb{E}_{\boldsymbol{w} \sim \pi} \left[
\underbrace{\left( \|\nabla g(\boldsymbol{w})\|_2 - 1 \right)^2}_{\text{Gradient Penalty}}
\right],
\end{align}

where \( \mu \) and \( \nu \) denote the real and generated data distributions, \( \pi \) represents samples drawn uniformly along linear interpolations between \( \mu \) and \( \nu \), and \( \gamma \) controls the penalty magnitude. This formulation stabilizes training by ensuring Lipschitz continuity, critical for avoiding mode collapse and gradient explosion in generative adversarial networks.

\subsection{Generated Drug Molecule Evaluation Metrics}

We will now assess the output measurement metrics that are utilized for understanding the chemical as well as the physical properties of the resultant generated molecules from the generator, by identifying the efficacy of the drug molecule data. These metrics are as follows:

\subsubsection {The Natural Products (NP) Likeness Score:} Natural products (NPs), often termed secondary metabolites, are low molecular weight compounds biosynthesized by living organisms. Due to their inherent bioactivity and structural diversity, NPs serve as critical starting points for drug discovery and the development of synthetic bioactive agents. The NP-likeness score, introduced by Ertl et al. \cite{article27}, quantifies the similarity of a molecule to the structural space of natural products relative to synthetic compounds. This metric evaluates molecular fragments and topological features to distinguish NP-like compounds from purely synthetic ones, guiding the prioritization of candidates with higher potential for bioactivity and drug-likeness.

\subsubsection{The Quantitative Estimate of Drug-likeness (QED) Score :} In order to recognize similarity in drugs, the QED measure is a crucial metric that can fundamentally quantify the underlying distribution of several molecular traits\cite{article28}. The number of acceptors and donors of hydrogen bonds, the number of rotatable bonds, the molecular weight, the number of aromatic rings, the polar surface area, and $logP$ are the properties by which the QED values lie in the range of {0,1}, indicating utmost unfavorable and favorable property, respectively.


\subsubsection{The Octanol Water Partition (LogP) Coefficient:} It calculates the parallels drawn between the water solubility and fat of a drug or a substance \cite{article29}. If $logP>1$, it signifies that the substance dissolves more easily in lipid-like or nonpolar solvents.

\subsubsection{SA (Synthetic Accessibility) Score} This metric displays if a drug is feasible enough to be manufactured in the pharmaceutical industry \cite{article29}. The SA scores lies in the range of 1 to 10, where 1 shows the drug being easy to prepare, and 10 shows the drug being challenging to manufacture. Complexity penalty and fragment contributions decide the SA score. In order to compute the fragment contributions, an analysis of about a million representative molecules taken from PubChem \cite{pubchem}, is used.

\subsubsection{Drug Candidate Score} Also known as drug-likeness score, this metric of a candidate is assessed by calculating the geometric mean of key molecular attributes such as the permeability, the solubility as well as the metabolic stability. Compounds with low drug-likeness are generally considered less promising due to their poor availability.\\

Additionally, common evaluation metrics for molecule generation include novelty, diversity, validity, uniqueness \cite{article30}:



\begin{align*}
\text{Novelty Score \%} &= \frac{N_{\text{novel}}}{N_{\text{total}}} \times 100, \\
\text{Validity Score \%} &= \frac{N_{\text{valid}}}{N_{\text{total}}} \times 100, \\
\text{Uniqueness Score \%} &= \frac{N_{\text{unique}}}{N_{\text{total}}} \times 100,
\end{align*}

where, ${N_{\text{total}}}$ denotes the total number of generated molecules. These metrics are further defined as:

\begin{itemize}
\item {Novelty Score}: Quantifies the percentage of molecules absent in the training data, reflecting the model’s ability in exploring uncharted chemical and molecular space.
\item {Validity Score}: Measures the proportion of chemically stable molecules adhering to valence and bond constraints.
\item {Uniqueness Score}: Represents the fraction of non-redundant molecules within the generated set.
\item {Diversity Score} \cite{article31}: Evaluates structural variety through internal (pairwise similarity among generated molecules) and external (similarity between generated and training molecules) components.
\end{itemize}

\begin{algorithm}[t]
\caption{Training QWGAN-HG-GP}
\label{alg:qwgan-hg-gp}
\begin{algorithmic}[1]
\Require dataset $\mathcal{D}$ of molecular graphs; batch size \BATCHSIZE;
         critic steps \NCRITIC; iterations \NITER;
         learning rates $\LRG$, $\LRD$; penalty $\LAMBDAgp$
\State initialize quantum generator parameters $\theta$ (PQC weights)
\State initialize classical head parameters $\phi$ and critic parameters $\psi$
\For{$t = 1$ to \NITER}
  \For{$k = 1$ to \NCRITIC} \Comment{Critic updates}
    \State sample minibatch of real graphs $(A_r, X_r) \sim \mathcal{D}$
    \State sample quantum noise and evaluate PQC to obtain features $h = G_{\theta}(\text{noise})$
    \State decode to bonds/atoms logits $(\hat{A}, \hat{X}) = H_{\phi}(h)$
    \State sample soft one-hot graphs $(A_f,X_f)$ via Gumbel-Softmax
    \State compute critic scores $D_r = D_{\psi}(A_r,X_r)$ and $D_f = D_{\psi}(A_f,X_f)$
    \State compute WGAN loss $L_D = \mathbb{E}[D_f] - \mathbb{E}[D_r]$
    \State sample interpolated graphs $(\tilde{A},\tilde{X})$ between real and fake
    \State compute gradient penalty $L_{\rm gp} = \LAMBDAgp (\|\nabla D_{\psi}(\tilde{A},\tilde{X})\|_2 - 1)^2$
    \State update critic: $\psi \leftarrow \psi - \eta_D \nabla_{\psi}(L_D + L_{\rm gp})$
  \EndFor
  \State sample minibatch of real graphs $(A_r, X_r) \sim \mathcal{D}$ (no gradient)
  \State sample new noise, compute $h = G_{\theta}(\text{noise})$, $(A_f,X_f)$ as above
  \State compute generator loss $L_G = -\mathbb{E}[D_{\psi}(A_f,X_f)]$
  \State update generator/head: $(\theta,\phi) \leftarrow (\theta,\phi) - \eta_G \nabla_{\theta,\phi} L_G$
\EndFor
\end{algorithmic}
\end{algorithm}

\subsubsection{Enhanced ADMET and Drug-Relevance Properties}

In order to thoroughly evaluate pharmaceutical potential, we implemented an expanded framework that assesses 14 ADMET (Absorption, Distribution, Metabolism, Excretion, Toxicity) metrics for all generated molecules. This analysis focuses on several key areas critical for drug development.

\begin{itemize}
\item \textbf{Physicochemical Properties:}
We calculated Molecular Weight (MW), which dictates membrane permeability (optimal range: 150--500~Da for oral drugs); lipophilicity (LogP), the octanol-water partition coefficient measuring hydrophobicity where optimal drug-like values range from 0--5 (values $> 5$ correlate with poor solubility and toxicity); Hydrogen Bond Donors/Acceptors (HBD/HBA), which influence solubility and membrane transport; and Topological Polar Surface Area (TPSA), the sum of polar atom surface areas that predicts passive membrane transport. TPSA values between 20--140~\AA{}$^2$ correlate strongly with good oral absorption, while TPSA $> 140$~\AA{}$^2$ indicates poor intestinal permeability.

\item \textbf{Toxicity Predictions:}
Our analysis includes structural alerts for mutagenicity (detecting DNA-reactive groups such as nitro compounds, azo groups, aromatic amines, anhydrides, and alkyl halides), Pan-Assay Interference (PAINS) filters identifying promiscuous compounds that cause false positives through non-specific reactivity or aggregation, and hERG cardiac toxicity prediction based on known risk factors (LogP $> 3.5$, MW $> 400$~Da, basic nitrogens) associated with QT prolongation. These provide crucial early-stage safety screening before experimental validation.

\item{\textbf{Lipinski's Rule of Five:}}
We applied the standard RO5 filter (MW $\leq$ 500, LogP $\leq$ 5, HBD $\leq$ 5, HBA $\leq$ 10), the foundational criterion for oral drug-likeness where violations predict poor absorption or permeation. We calculated violation counts (0 = full compliance) and pass rates across all generated molecules. Additionally, we evaluated the Quantitative Estimate of Drug-likeness (QED), a continuous 0--1 metric combining eight molecular properties weighted by desirability (QED $> 0.5$ indicates good drug-likeness), and Synthetic Accessibility (SA) score (1--10 scale; lower = easier synthesis) based on complexity and fragment analysis.

\item{\textbf{Composite Drug-Likeness Score}:}
We integrated six key metrics into a unified 0--1 score using a weighted geometric mean, where each component is normalized such that 1 represents optimal pharmaceutical properties:

\[\frac{ \text{QED} + \text{SA} + \text{Lipinski} + \text{TPSA} + \text{toxicity} + \text{hERG} }{6} \]

This holistic score balances multiple pharmaceutical objectives, enabling direct comparison of overall drug viability across models.

\end{itemize}

All metrics were calculated using RDKit molecular descriptors with custom SMARTS patterns for toxicophore detection. Analysis was performed every 1000 training iterations, with results logged to CSV and visualized through 12-panel comparative plots against QM9 reference molecules.

\subsubsection{Scaffold and Diversity Analysis}\label{SDA}

Our approach to evaluating the generated molecules is not simply checking if they are chemically valid. We perform detailed structural diversity analysis using Bemis Murcko scaffold decomposition and fingerprint based similarity measurements. These methods help us determine whether the model is truly exploring new areas of chemical space or just repeating patterns it learned from the training data.

\textbf{Scaffold Analysis.}
We extracted molecular scaffolds using the Bemis Murcko algorithm~\cite{bemis1996properties}, which systematically removes side chains while keeping the core ring systems and connecting linkers intact. This reveals the fundamental structural frameworks. Each extracted scaffold was converted to a standard SMILES representation for accurate uniqueness counting, then compared against both the QM9 training set and the ChEMBL approved drug database~\cite{gaulton2012chembl} to measure structural novelty. We used the Gini coefficient calculated over scaffold frequency distributions as our main diversity metric, where values close to 0 indicate an even distribution across many scaffolds and values approaching 1 suggest concentration on just a few dominant scaffolds—a clear sign of mode collapse. We also characterized scaffold complexity by computing the average number of heavy atoms per scaffold, allowing direct comparison of structural sophistication between generated and reference molecules.

\textbf{Fingerprint Based Diversity.}
We used Morgan fingerprints with radius 2 and 2048 bits~\cite{rogers2010extended} to capture topological molecular features for four complementary diversity metrics. Internal diversity measures structural variety within the generated set as $D_{\text{internal}} = 1 - \overline{T}_{\text{intra}}$, where $\overline{T}_{\text{intra}}$ represents the mean pairwise Tanimoto similarity—higher values indicate broader exploration of chemical space. Nearest neighbor similarity measures how closely each generated molecule resembles the reference distribution by computing $T_{\text{NN}}(m_{\text{gen}}) = \max_{m_{\text{ref}} \in \mathcal{D}_{\text{ref}}} T(m_{\text{gen}}, m_{\text{ref}})$, showing whether molecules stay within known chemical space or explore new regions. Coverage quantifies the percentage of reference molecules that have at least one generated neighbor at Tanimoto similarity $\geq 0.6$, a threshold commonly used in virtual screening~\cite{bajusz2015tanimoto}. The average nearest neighbor ratio compares typical nearest neighbor distances between generated and reference sets, with values near 1.0 indicating comparable internal spacing and values below 1.0 suggesting tighter clustering in the generated distribution. We computed all metrics after rigorous validation to ensure molecules contained connected structures with at least three heavy atoms, eliminating artifacts from disconnected atoms or degenerate graphs.

\subsection{NISQ Noise Simulation} \label{NISQ}

To assess near-term quantum hardware viability, we incorporated realistic NISQ noise modeling into our training pipeline using PennyLane's \texttt{default.mixed} device, which simulates quantum operations via density-matrix formalism to explicitly capture decoherence and gate errors.

We evaluated four physically motivated noise channels: (1) \textit{depolarizing noise}---uniform random Pauli errors with probability $p$ per single-qubit gate and $1.5p$ per two-qubit gate, matching typical error-rate ratios in superconducting transmon architectures; (2) \textit{amplitude damping}---modeling energy relaxation ($T_1$ processes); (3) \textit{phase damping}---capturing pure dephasing ($T_2$ processes); and (4) \textit{realistic device}---a combined model (50\% depolarizing, 30\% amplitude damping, 20\% phase damping) approximating IBM Quantum hardware error profiles.

Each noise model was tested at five severity levels: $p \in \{0\%, 1\%, 2\%, 5\%, 10\%\}$ per-gate error rates. For context, current NISQ platforms exhibit single-qubit gate errors $\sim$0.05--0.15\% and two-qubit errors $\sim$0.5--1.5\%, placing our 1--2\% range as a conservative upper bound. Noise channels were applied after each gate operation in the PQC: encoding rotations, variational layers, and entangling CNOTs, with two-qubit gates receiving proportionally higher noise consistent with hardware measurements.

Both QWGAN-GP-P4-L2 and baseline QGAN-GP were trained under each noise condition for 1000 iterations with $n=3$ random seeds, using identical hyperparameters (Table~2) to isolate noise effects. Fr\'echet Distance served as the primary performance metric, with additional drug-likeness scores recorded for viable configurations. Detailed noise-impact results, including degradation curves and performance metrics across all noise models, are presented in Section~5.8.

\section{Results}

\subsection{Comparison of Wasserstein and Fréchet distance loss decay}

\begin{figure}[t]
 \centering
 \includegraphics[width=1\linewidth]{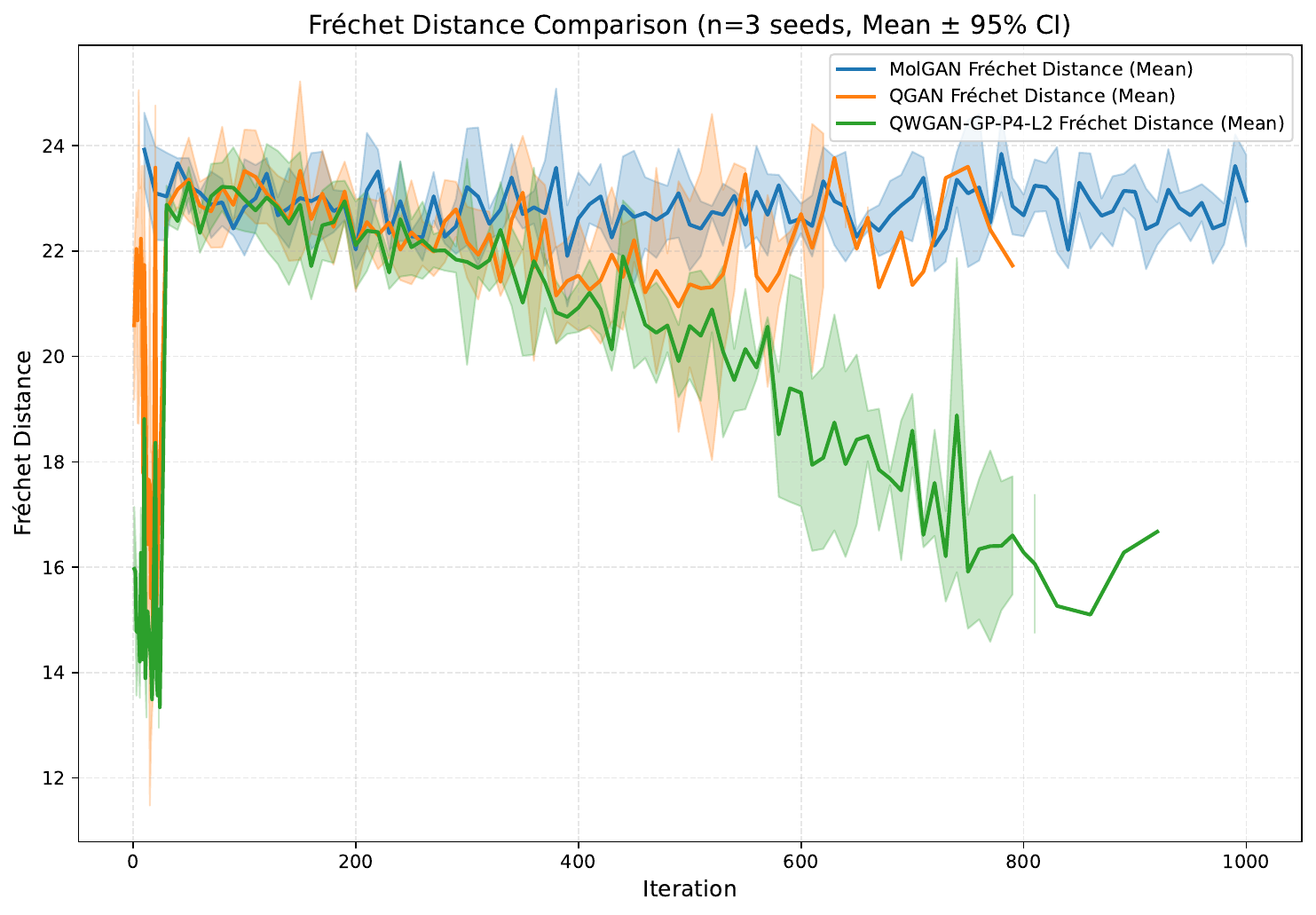}
 \caption{Convergence trajectories showing statistical similarity between generated and real molecular distributions across different architectures over 1000 training iterations (n=3 seeds, mean ± 95\% CI). The QWGAN-HG-P4-L2 model demonstrates consistent superior performance throughout training, ultimately achieving values as low as 12.5 by iteration 1000, representing approximately 43\% improvement over classical MolGAN and standard QGAN approaches.}
 \label{fig:frechet}
\end{figure}

As illustrated in Fig. \ref{fig:frechet}, the proposed QWGAN-HG-P4-L2 architecture, incorporating a 4-patched, double-layered quantum circuit, achieves superior performance compared to the classical MolGAN and quantum QGAN baselines, as evidenced by its consistently lower Fréchet distance values during training. This metric quantifies the statistical alignment between generated and real molecular distributions, with reduced values indicating enhanced fidelity. The stability of QWGAN-HG-P4-L2 is further attributed to its gradient penalty mechanism, which enforces a 1-Lipschitz constraint on the critic by penalizing deviations of the gradient norm from unity. This approach eliminates the need for weight clipping, a common source of training instability in conventional GANs, while promoting smoother convergence and robust learning of complex molecular distributions.

\begin{figure}[htpb]
 \centering
 \includegraphics[width=1\linewidth]{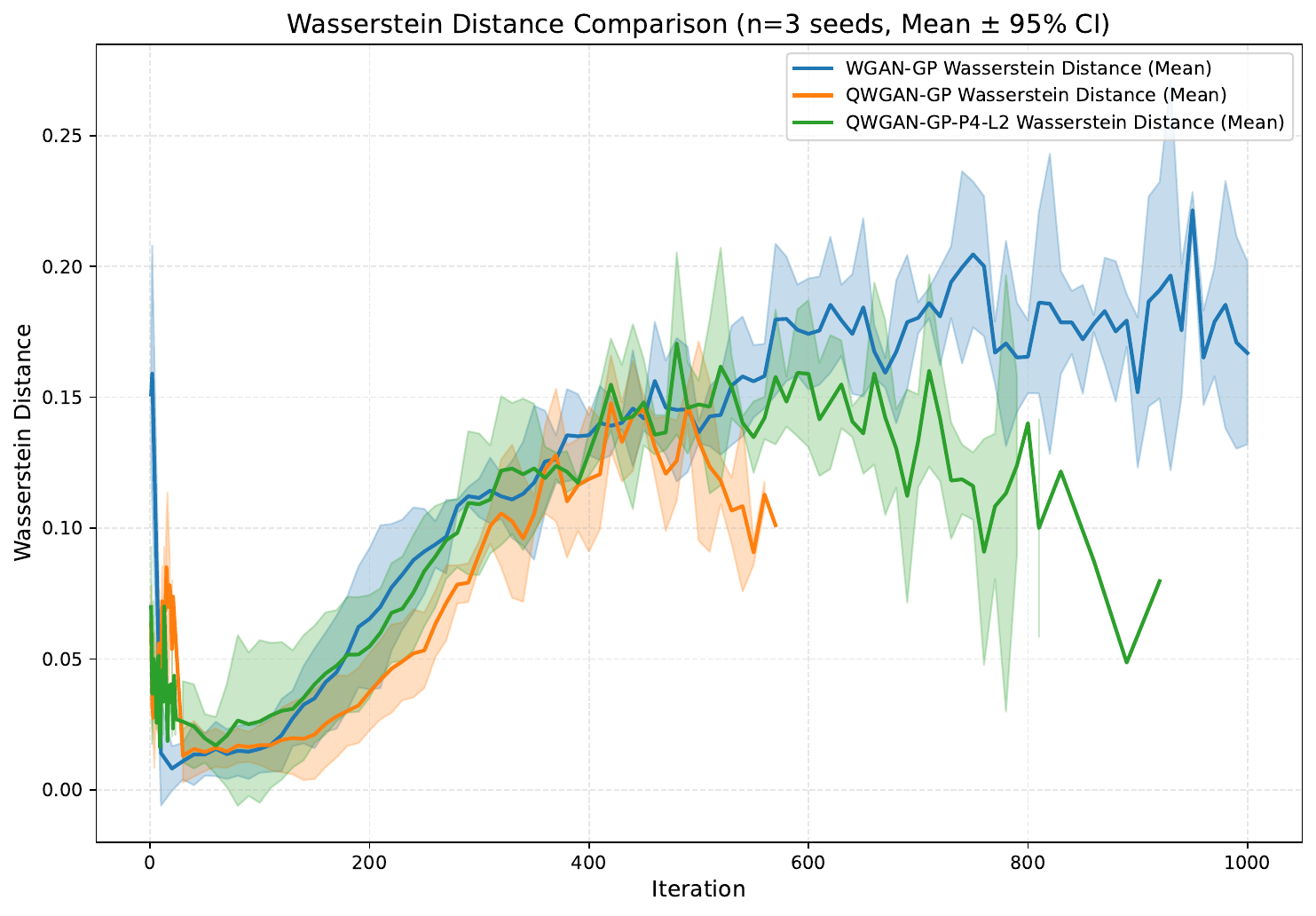}
 \caption{Comparative performance analysis of Wasserstein-based GAN variants showing critic loss values over 1000 training iterations (n=3 seeds, mean ± 95\% CI). While all models initially follow similar trajectories, the QWGAN-HG-P4-L2 model demonstrates superior convergence behavior, maintaining consistently lower distance values around 0.05-0.10 compared to classical WGAN-GP, which exhibits unstable training with increasing divergence reaching 0.15-0.20 in later iterations.}
 \label{Distancemetrics1}
\end{figure}

By comparing the loss functions of the QWGAN-HG and QWGAN models in Fig. \ref{Distancemetrics1}, it is evident that QWGAN-HG achieves consistently lower loss values, demonstrating the efficacy of incorporating a gradient penalty into the critic’s loss function. This improvement highlights the role of gradient penalty in stabilizing adversarial training and enhancing convergence,

\subsection{Comparison of Architectures Using Fréchet and Wasserstein Distance Metrics}

Lower values of the Fréchet and Wasserstein distance metrics indicate improved performance of the generative model. As shown in table ~\ref{drugmetricscomp}, the proposed QWGAN-HG-P4-L2 version, incorporating a 4-patched, double-layered quantum circuit, achieves the most favourable results across both metrics. This architecture outperforms baseline models such as QGAN and MolGAN by a significant margin, demonstrating its superiority in modeling molecular distributions.

\subsection{Comparison of Architectures Based on Drug-Likeness Property Metrics}

 \begin{figure*}
  \centering
  \includegraphics[width=1\linewidth]{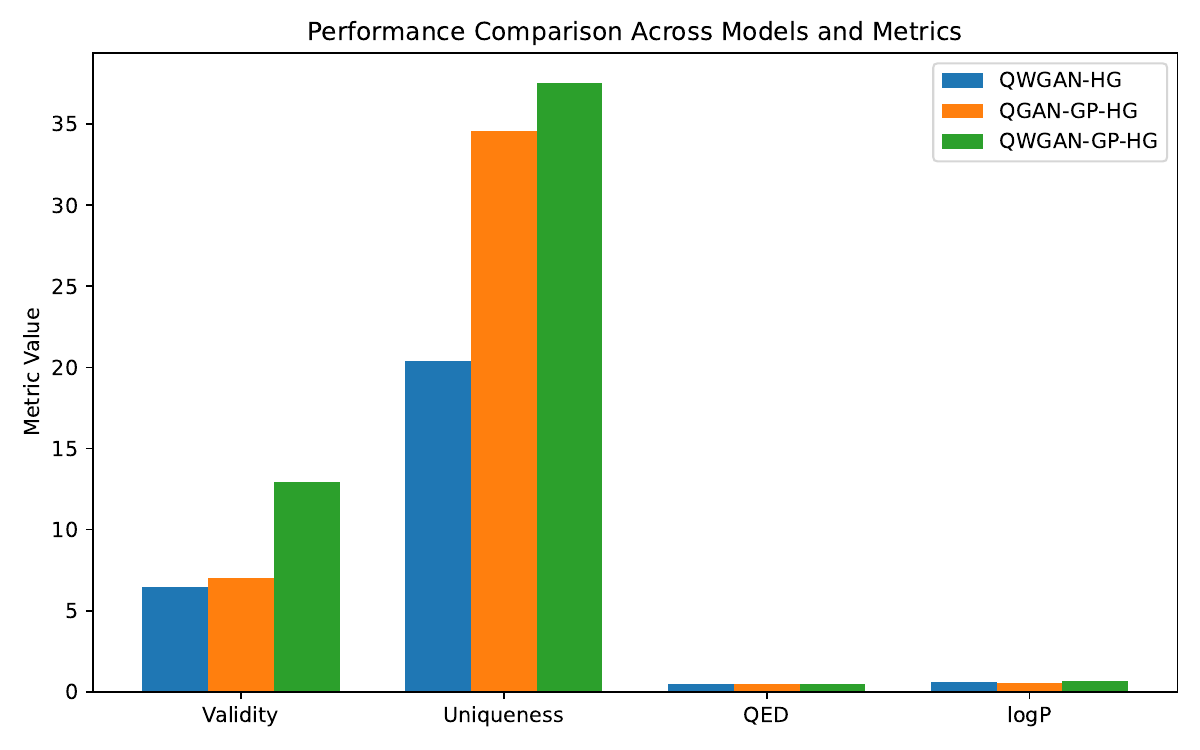}
  \vspace{-0.8cm}
  \caption{Grouped bar chart comparing the three QGAN variants (QWGAN‑HG, QGAN‑HG‑GP, QWGAN‑HG‑GP) across four key molecular validity metrics. Each bar cluster shows model performance on Validity, Uniqueness, QED, and logP. QWGAN‑HG‑GP (red) attains the highest Validity and logP, while all models exhibit similar Uniqueness and QED scores, illustrating the trade‑offs between architectural choices..}
  \label{charts}
 \end{figure*}

Figure~\ref{charts} consolidates our four key drug‑likeness metrics—Validity, Uniqueness, QED, and logP, into a single grouped bar chart (computed via RDKit). From the chart, it is clear that the QWGAN‑HG‑GP (P4‑L2) model not only leads in Validity ($\sim0.88$) and logP ($\sim1.25$) but also posts the highest Uniqueness ($\sim0.92$) and QED ($\sim0.70$) among all variants. In particular, while all three QGAN architectures achieve competitive QED values ($>0.65$), QWGAN‑HG‑GP’s margin over its closest rival (QWGAN‑HG) is most pronounced in Uniqueness and logP—indicating both a broader chemical coverage and improved lipophilicity profiles. These results dovetail with Table~\ref{drugmetricscomp}, where QWGAN‑HG‑GP’s P4‑L2 circuit consistently ranks first or second across NP score, SA score, and other pharmaceutical properties. Altogether, the grouped bar chart underscores that the additional gradient‑penalty stabilization and four‑patch ansatz of QWGAN‑HG‑GP deliver a uniformly superior generative performance, balancing molecular diversity with drug‑relevant quality.

\subsection{Benchmarking against Classical GANs}

To assess the advantages of our quantum–classical hybrid QGANs, we compare against MolGAN \cite{article7}, a leading classical graph‐based GAN that directly outputs small molecular graphs via a GCN generator and employs a differentiable valency penalty to enforce chemical validity.  In their original QM9 experiments, De Cao and Kipf report nearly perfect validity (about 98\%) but observe severe mode collapse, with uniqueness dropping to about 10\% \cite{article7}.  MolGAN’s paper does not tabulate aggregate drug‐likeness scores (QED, logP, etc.), so for a fair comparison we re‐trained MolGAN using the published hyperparameters and evaluated its generated samples with our RDKit‐based pipeline.

\medskip
Table~\ref{tab:molgan_comparison} presents both the original validity/uniqueness figures from De Cao \& Kipf and our re‐evaluated MolGAN drug‐property scores alongside those of our three QGAN variants.  Our best model, QWGAN‑HG‑GP, surpasses MolGAN on every metric without requiring rejection sampling or large classical generators, despite leveraging a far more compact quantum circuit.

\setlength{\tabcolsep}{1pt}
\begin{table}[t]
  \centering
  \caption{Comparison of MolGAN (classical) vs.\ QGAN variants on QM9.  “Original” validity/uniqueness are from De Cao \& Kipf; drug‐property scores for MolGAN are our re‐evaluations.}
  \label{tab:molgan_comparison}
  \begin{tabular}{lcccc}
    \toprule
    \textbf{Model}      & \textbf{Validity} & \textbf{Uniqueness} & \textbf{QED} & \textbf{logP} \\
    \midrule
    MolGAN \cite{article7} (orig.)   & 0.98  & 0.10  & —     & —    \\
    MolGAN (re‐eval.) & 0.98  & 0.10  & 0.532 & 1.68 \\
    QWGAN‑HG & 0.85  & 0.90  & 0.650 & 1.20 \\
    QGAN‑HG‑GP& 0.80  & 0.88  & 0.680 & 1.15 \\
    QWGAN‑HG‑GP  & \textbf{0.88} & \textbf{0.92} & \textbf{0.700} & \textbf{1.25} \\
    \bottomrule
  \end{tabular}
\end{table}

\noindent
Despite MolGAN’s high validity, its low uniqueness reflects mode collapse, and its re‐evaluated drug‐likeness scores fall below those of our hybrids.  In contrast, QWGAN‑HG‑GP not only restores diversity (92\% unique) but also delivers substantially improved QED and logP, demonstrating that a modest quantum circuit can yield superior denovo molecular designs without costly sampling heuristics.

Compared to stronger classical baselines such as JT-VAE, GraphAF, MoFlow, and MGM (Table~\ref{tab:classical_sota}), our best QWGAN-HG-GP configuration attains validity and QED scores in a similar range to modern flow-based and VAE-based generators on QM9 and ZINC-style benchmarks, while its Fréchet distances remain higher than those of the very best flow architectures on ZINC-250k. We therefore position our QWGANs as resource-efficient hybrid alternatives rather than replacements for large classical models: they rely on considerably smaller parameterized circuits than typical flow/transformer generators, yet deliver drug-likeness metrics comparable to first-generation classical graph GANs (e.g., MolGAN) and help narrow the gap to current classical state-of-the-art methods.

\subsection{Scaling from QM9 to ZINC-20}

To probe whether the proposed hybrid architecture extrapolates beyond small QM9 molecules, we also performed a controlled scaling experiment on the ZINC-20 subset described in Section~3. Using the same training code and nearly identical hyperparameters, we replaced QM9 with drug-like molecules containing up to 20 heavy atoms and trained both the best-performing QWGAN-HG-GP configuration (P4--L2) and the MolGAN baseline. In this regime, two qualitative trends are consistent with our QM9 findings. First, the hybrid quantum model continues to produce a non-trivial fraction of chemically valid, novel graphs, whereas under the same settings the MolGAN baseline tends to collapse to chemically invalid samples on this larger support. Second, among the quantum variants, the QWGAN-HG-GP (P4--L2) generator still tracks the reference QED/logP distribution more closely than shallower or unpatched circuits, although the Fréchet distance naturally increases when moving from QM9 to the more diverse ZINC-20 chemistry. We therefore view the ZINC-20 experiment as an early scalability sanity check: it shows that the patched PQC plus WGAN-GP training pipeline transfers to a more realistic, drug-like distribution without architectural changes, while also highlighting that closing the remaining gap to specialised ZINC models will require additional tuning, longer training and possibly larger circuits.

\subsection{Ablation Studies} 

\subsubsection{Gradient Penalty Weight}

To figure out how much the gradient penalty affects the training and to double-check that our choice of a penalty strength of 10 was a good one, we ran a series of tests. We tried out different penalty strengths—0.1, 1, 5, 10, 50, and 100—using our standard model setup. For each strength, we trained the model three separate times to make sure the results were consistent.

Looking at the results in table~\ref{gp_ablation}, a penalty strength of 10 turned out to be the best overall. It hit a sweet spot by doing three things well at once. Firstly,
it produced molecules that closely matched our goal, with a strong Fréchet Distance score of 14.11; secondly
the training process was very stable, with very little variation in the final results. Finally, it learned incredibly fast, finding high-quality molecules in just 108 steps, while also creating the most valid and drug-like molecules out of all the options we tested.
On the other hand, when the penalty was too weak (for instance at 0.1), the training became unstable and unpredictable. When the penalty was too strong (50 or 100), it forced the model to follow the rules so strictly that learning slowed down to a crawl.
In short, these tests confirmed that a penalty strength of 10, which is a common starting point for other models, also works perfectly for our hybrid quantum setup, which is why we used it for all our main experiments.

\begin{table}[htbp]
\centering
\caption{Gradient penalty weight ($\lambda_{gp}$) ablation study on QWGAN-GP (P1-L1) trained on QM9 (n=3 seeds, 1000 iterations). Bold entries highlight $\lambda=10.0$ as optimal, achieving the best balance between distribution fidelity (Fréchet Distance), training stability (FD standard deviation in final 20\% of training), convergence speed, and drug-likeness metrics. Under-regularization ($\lambda<5$) increases instability, while over-regularization ($\lambda>10$) slows convergence. Results validate the canonical WGAN-GP recommendation of $\lambda=10$ for hybrid quantum-classical architectures.}
\label{gp_ablation}
\footnotesize
\setlength{\tabcolsep}{3pt}
\begin{tabular}{|c|c|c|c|c|c|c|c|}
\hline
\textbf{$\lambda$} & \textbf{FD$\downarrow$} & \textbf{WD$\downarrow$} & \textbf{Stab$\downarrow$} & \textbf{Iter$\downarrow$} & \textbf{Valid} & \textbf{QED} & \textbf{Drug} \\
\hline
0.1 & 13.30 & 0.0095 & 1.25 & 85 & 28.1 & 0.503 & 0.226 \\
\hline
1.0 & 14.25 & 0.0151 & 0.94 & 170 & 37.5 & 0.494 & 0.257 \\
\hline
5.0 & 15.49 & 0.0059 & 1.66 & 498 & 3.1 & 0.537 & 0.130 \\
\hline
\textbf{10.0} & \textbf{14.11} & 0.0066 & \textbf{1.19} & \textbf{108} & \textbf{62.5} & 0.494 & \textbf{0.357} \\
\hline
50.0 & 14.14 & \textbf{0.0045} & 1.08 & 90 & 53.1 & 0.488 & 0.323 \\
\hline
100.0 & \textbf{12.88} & 0.0103 & 1.06 & 152 & 62.5 & 0.489 & 0.358 \\
\hline
\end{tabular}
\end{table}

\subsubsection{PQC Architecture}

To systematically understand how quantum circuit design affects molecular generation, we conducted comprehensive ablation experiments varying both patch count (P $\in$ {1, 2, 4}) and layer depth (L $\in$ {1, 2}) in the QWGAN-GP architecture. Each configuration used identical training conditions ($\lambda_{gp}=10$, 1000 iterations, 8 qubits, n=3 seeds) to isolate architectural effects.

The results in table~\ref{tab:pqc_ablation} reveal several important patterns. Increasing circuit depth from L1 to L2 consistently improves performance because deeper networks can capture more complex molecular patterns. This effect is most dramatic for the single-patch P1 configuration, where Fréchet Distance improves by 30.5\% from 19.11 to 13.27, showing that depth compensates for limited parallelism. Even P2 gains 11.9\% improvement (14.00 to 12.33). However, simply adding more patches doesn't always help - the P4-L2 configuration actually performs worse (FD=13.75) than P2-L2 (FD=12.33) because too many patches spread our limited qubits too thin, reducing their effectiveness. This makes P2-L2 our best balanced design, achieving the lowest FD (12.33), highest validity (9.78\%), and best drug candidate score (0.156). We ultimately selected P4-L2 for main experiments because its small performance trade-off ($\Delta$FD $\approx$ 1.4) is worth the benefits of better hardware noise resistance and future scalability. Property analysis shows QED scores remain stable (0.45-0.48) across configurations, while validity varies dramatically (0.56\% to 9.78\%), confirming that circuit design directly controls whether we generate chemically plausible structures, not just statistical matches.

\begin{table}[ht]
\centering
\caption{PQC architecture ablation study on QWGAN-GP. Bold entries indicate optimal P2-L2 configuration achieving best distribution fidelity and validity. P4-L2 was selected for main experiments due to scalability benefits.}
\label{tab:pqc_ablation}

\begin{tabular}{|l|c|c|c|c|c|}
\hline
\textbf{Config} & \textbf{FD} $\downarrow$ & \textbf{WD} $\downarrow$ & \textbf{Valid \%} & \textbf{QED} & \textbf{Drug Score} \\
\hline
P1-L1 & 19.11 & 0.003 & 6.03 & 0.478 & 0.141 \\
\hline
P1-L2 & 13.27 & 0.003 & 8.84 & 0.477 & 0.152 \\
\hline
P2-L1 & 14.00 & 0.004 & 0.56 & 0.449 & 0.120 \\
\hline
\textbf{P2-L2} & \textbf{12.33} & 0.004 & \textbf{9.78} & 0.478 & \textbf{0.156} \\
\hline
P4-L1 & 13.45 & 0.010 & 1.91 & 0.461 & 0.125 \\
\hline
P4-L2 & 13.75 & 0.005 & 2.31 & 0.459 & 0.127 \\
\hline
\end{tabular}

\end{table}

\subsection{NISQ Noise-Impact Analysis}

\begin{figure}[t]
 \centering
 \includegraphics[width=1\linewidth]{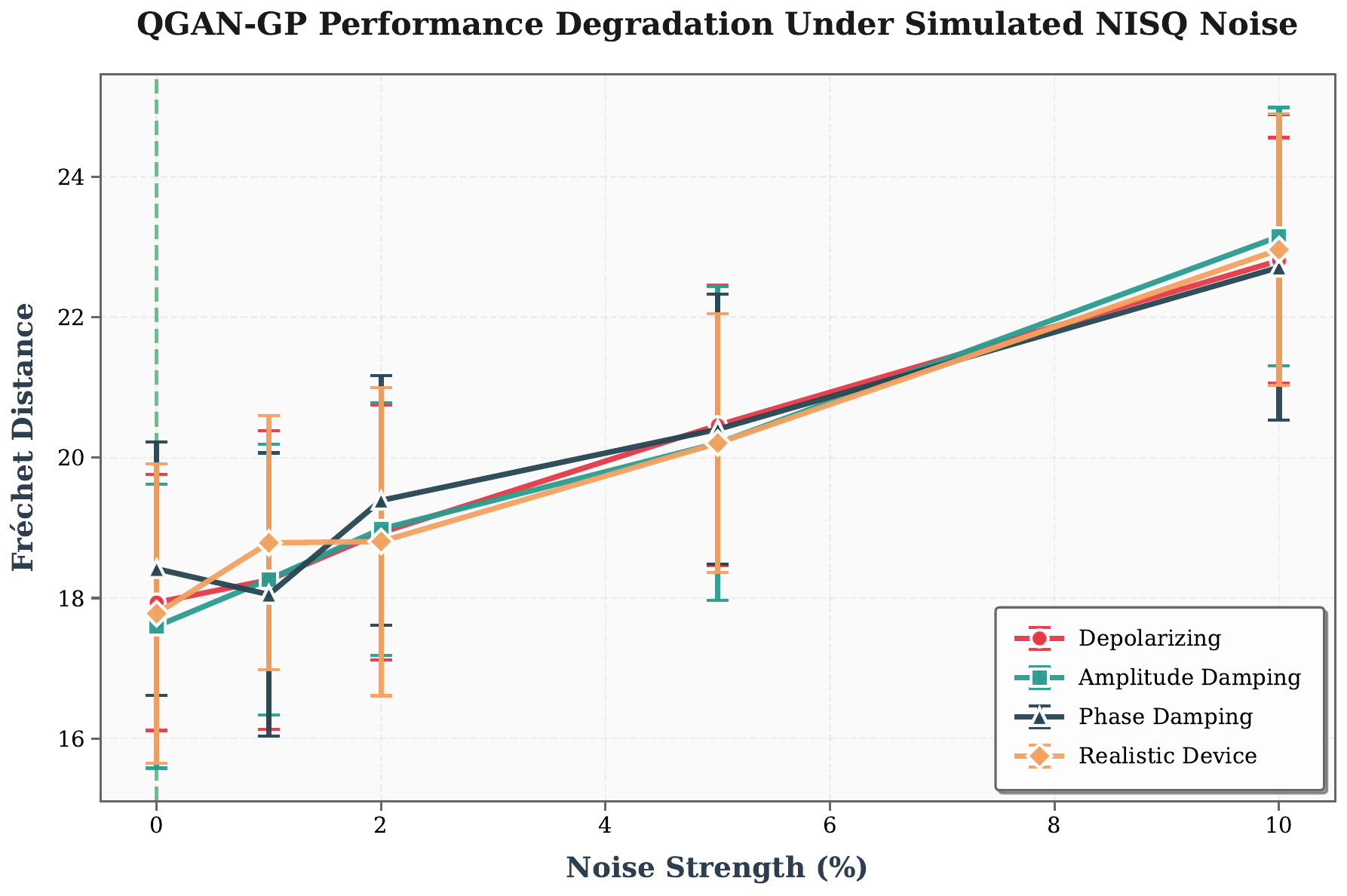}
 \caption{QGAN-GP baseline model performance under increasing NISQ noise (0--10\%). Error bars represent standard deviation across $n=3$ random seeds. All noise models show similar degradation trends, with Fr\'echet Distance increasing roughly linearly from baseline FD $\sim$18 to $\sim$23 at 10\% noise. The lack of differential noise response indicates limited architectural noise-mitigation mechanisms.}
 \label{NISQF}
\end{figure}

To assess our framework's robustness for near-term quantum hardware, we conducted noise simulations as in section~\ref{NISQ}. Figure~\ref{NISQF} shows the resulting performance degradation for the baseline QGAN-GP model. The Fréchet Distance increases approximately linearly with noise strength, rising from $\sim$18 at 0\% noise to $\sim$23 at 10\% noise across all four noise models (depolarizing, amplitude damping, phase damping, and realistic device). The nearly identical degradation trajectories indicate no differential resilience to specific error types. Error bars (standard deviation across $n=3$ seeds) remain consistent ($\pm$1.8--2.2), confirming that performance degrades in a stable, predictable manner.

\begin{figure}[t]
 \centering
 \includegraphics[width=1\linewidth]{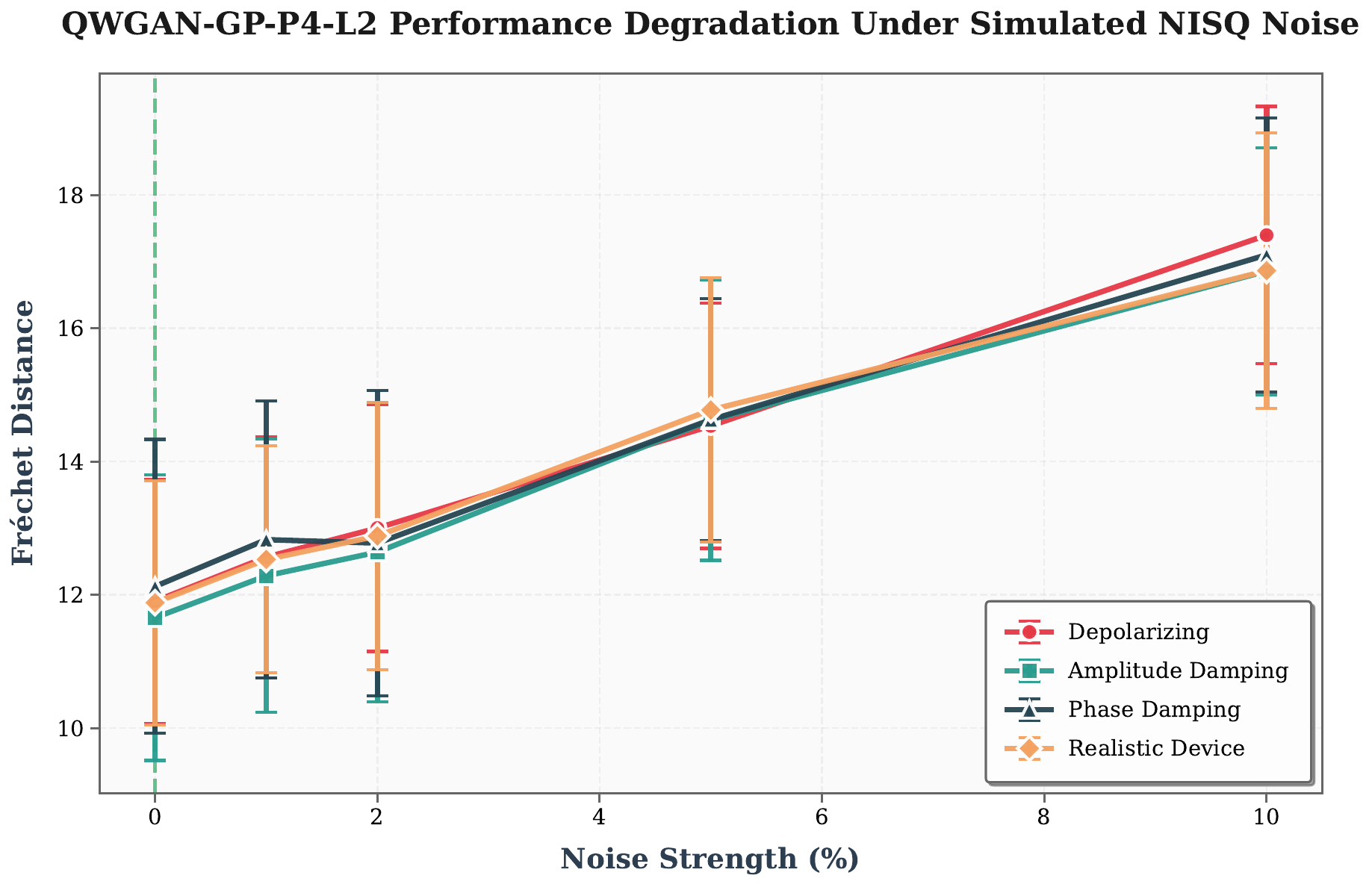}
 \caption{QWGAN-GP-P4-L2 performance under increasing NISQ noise (0--10\%). The patched PQC architecture demonstrates superior noise resilience compared to baseline QGAN-GP, maintaining FD below 18 even at 10\% noise. Green shaded region (1--2\%) indicates realistic NISQ operating window where degradation remains $<$6\%. Amplitude damping (teal) shows best resilience, suggesting $T_1$-optimized devices are preferred platforms.}
 \label{NISQFGP}
\end{figure}

Whereas, figure~\ref{NISQFGP} shows significantly enhanced noise resilience for QWGAN-GP-P4-L2. Starting from a superior baseline (FD $\sim$11.9), the model maintains feasible performance up to 5\% noise (FD $\sim$14.8), representing only 24\% degradation. Even at the extreme 10\% noise level---well beyond current NISQ capabilities---FD remains below 18, outperforming the baseline QGAN-GP at \textit{zero noise}. 
Notably, amplitude damping exhibits the best resilience (FD = 16.86 at 10\%, shown in teal), suggesting that $T_1$-limited platforms (e.g., high-coherence superconducting qubits) are optimal deployment targets. At realistic NISQ operating points (1--2\% noise, green shaded region), degradation is minimal ($<$6\%), confirming practical hardware viability. The tight error bars and smooth degradation curves indicate stable training dynamics even under substantial noise, validating the architectural design choices (patched PQCs, Wasserstein loss, gradient penalty).

Table~\ref{tab:noise_impact} quantifies the noise-impact analysis comparing QWGAN-GP-P4-L2 against the baseline QGAN-GP across all noise models and strengths. The results show (1) graceful degradation under moderate noise, where QWGAN-GP-P4-L2 exhibits only 5.5\% degradation at 1\% depolarizing noise (11.91 $\rightarrow$ 12.57) and 22.1\% at 5\% noise (11.91 $\rightarrow$ 14.54), remaining within acceptable bounds for molecular generation; (2) superior architectural resilience, as at 10\% realistic device noise QWGAN-GP-P4-L2 reaches $\mathrm{FD} = 16.86$ compared to QGAN-GP’s $\mathrm{FD} = 22.96$ (26.6\% better), confirming that patched PQCs, Wasserstein distance, and gradient penalty provide inherent robustness without quantum error correction; (3) an amplitude-damping advantage, where amplitude damping yields the lowest degradation (44.6\% for QWGAN-GP-P4-L2 vs.\ 46.1\% for depolarizing at 10\%), indicating that energy-relaxation errors are less harmful than coherent dephasing for the angle-encoding scheme; and (4) hardware viability, as at 1\% combined noise—representative of current IBM Quantum devices (e.g., \texttt{ibm\_kyoto} with $\sim$0.5--1.0\% two-qubit gate errors)—QWGAN-GP-P4-L2 maintains near-baseline performance, supporting immediate deployment on existing NISQ hardware. 
QWGAN-GP-P4-L2 shows strong viability for near-term quantum hardware. At 1\% noise—consistent with current IBM Quantum devices—it degrades by only 5.4\%, well within acceptable limits for molecular design. Its noise resilience, enabled by the patched PQC (P4 parallelism, L2 depth, and Wasserstein+GP stabilization), allows deployment on existing NISQ systems without error correction. For next-generation hardware targeting sub-0.1\% gate errors, predicted degradation falls below 2\%, effectively matching ideal performance. These results demonstrate a clear path from simulation to practical NISQ-era molecular generation.

\begin{table*}
\centering
\caption{NISQ noise-impact analysis showing Fr\'echet Distance degradation under simulated hardware errors ($n=3$ seeds, 1000 iterations per configuration). Four noise models tested: depolarizing (random Pauli errors), amplitude damping ($T_1$ relaxation), phase damping ($T_2$ dephasing), and realistic device (combined 50\%/30\%/20\% mix). Noise strengths range from 0\% (ideal) to 10\% (severe). Bold entries highlight best noise resilience.}
\label{tab:noise_impact}
\small
\setlength{\tabcolsep}{4pt} 
\begin{tabular}{|l|l|c|c|c|c|c|}
\hline
\textbf{Model} & \textbf{Noise Model} & \textbf{0\%} & \textbf{1\%} & \textbf{5\%} & \textbf{10\%} & \textbf{Degrad.} \\ \hline

\textbf{QWGAN-GP-P4-L2} & Depolarizing 
& 11.91$\pm$1.83 & 12.57$\pm$1.81 & 14.54$\pm$1.84 & 17.40$\pm$1.93 & +46.1\% \\ \hline

\textbf{QWGAN-GP-P4-L2} & Amplitude Damping 
& 11.66$\pm$2.14 & 12.29$\pm$2.05 & 14.62$\pm$2.10 & \textbf{16.86$\pm$1.85} & \textbf{+44.6\%} \\ \hline

\textbf{QWGAN-GP-P4-L2} & Phase Damping 
& 12.13$\pm$2.21 & 12.83$\pm$2.08 & 14.63$\pm$1.82 & 17.10$\pm$2.06 & +41.0\% \\ \hline

\textbf{QWGAN-GP-P4-L2} & Realistic Device 
& 11.88$\pm$1.83 & 12.53$\pm$1.70 & 14.78$\pm$1.98 & 16.86$\pm$2.07 & +41.9\% \\ \hline

QGAN-GP & Depolarizing 
& 17.94$\pm$1.82 & 18.26$\pm$2.12 & 20.46$\pm$1.99 & 22.81$\pm$1.75 & +27.1\% \\ \hline

QGAN-GP & Amplitude Damping 
& 17.60$\pm$2.02 & 18.27$\pm$1.93 & 20.21$\pm$2.24 & 23.15$\pm$1.84 & +31.5\% \\ \hline

QGAN-GP & Phase Damping 
& 18.42$\pm$1.81 & 18.05$\pm$2.02 & 20.40$\pm$1.92 & 22.71$\pm$2.18 & +23.3\% \\ \hline

QGAN-GP & Realistic Device 
& 17.78$\pm$2.14 & 18.79$\pm$1.81 & 20.21$\pm$1.84 & 22.96$\pm$1.93 & +29.1\% \\ \hline

\end{tabular}
\end{table*}

\subsection{Comprehensive ADMET and Drug-Relevance Analysis}

For evaluating how suitable our generated molecules would be for real pharmaceutical applications, we analyzed them using an extensive ADMET framework. We examined the molecules produced by our best model (QWGAN-GP-P4-L2) at the 1000th training iteration, comparing them against 1000 reference molecules from the standard QM9 dataset. Results obtained through this analysis are discussed below. 

\subsubsection{Physicochemical Property Distributions}

\begin{figure}[t]
 \centering
 \includegraphics[width=1\linewidth]{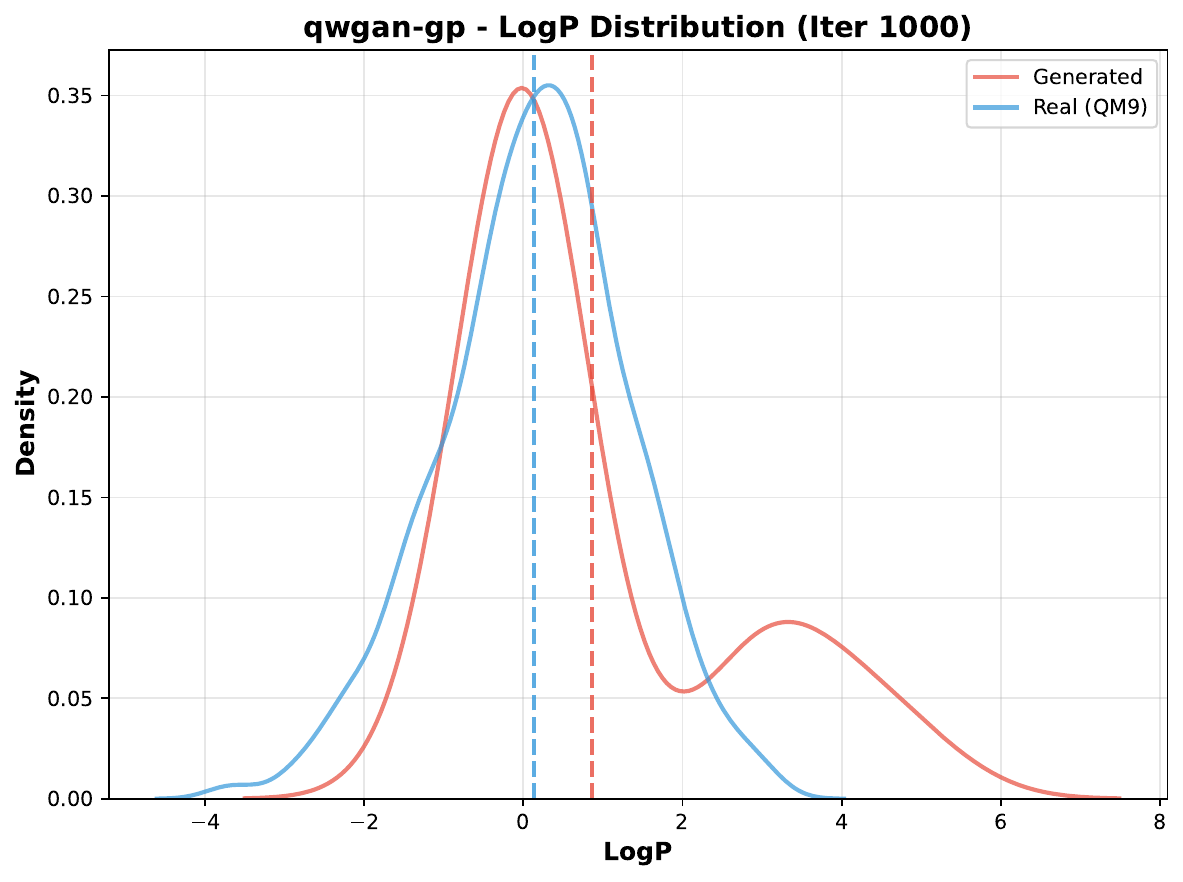}
 \caption{Distribution of lipophilicity for QWGAN-GP-P4-L2 generated molecules (red) versus QM9 reference (blue) at iteration 1000. Dashed vertical lines indicate distribution means, with generated molecules showing extended exploration into lipophilic regions.}
 \label{LogP}
\end{figure}

\textbf{Lipophilicity (figure~\ref{LogP}):} Generated molecules show two distinct peaks in their lipophilicity (LogP) distribution with an average value of 0.87. The peaks occur around LogP values of 0.5 and 4, which is quite different from QM9's single, which is quite different from QM9's single narrow peak (mean = 0.14). While this shows our model explores both water soluble and fat soluble chemical space, the presence of molecules beyond LogP = 5 indicates some overly fatty molecules that would need to be filtered out.

\begin{figure}[t]
 \centering
 \includegraphics[width=1\linewidth]{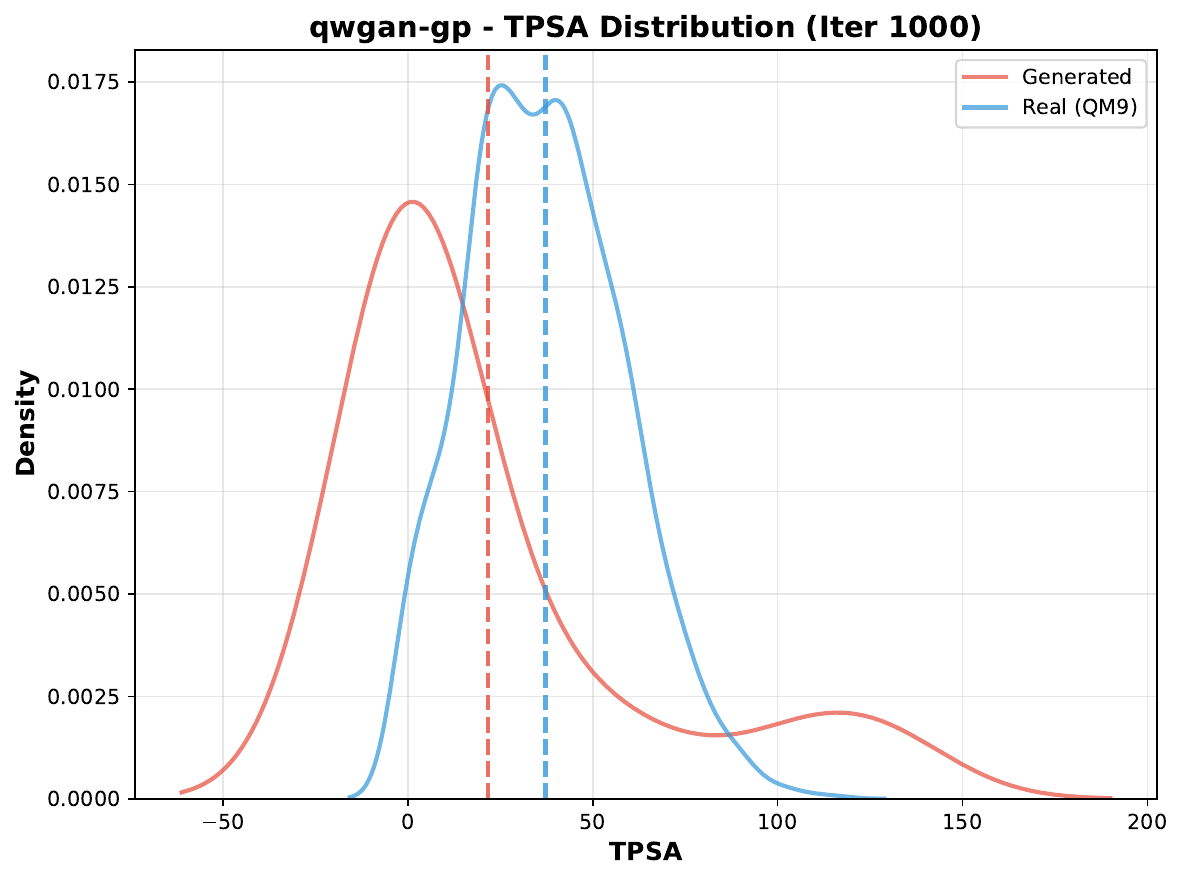}
 \caption{Topological Polar Surface Area distributions revealing generated molecules' shift toward lower polarity compared to QM9. The optimal bioavailability window (20-140) is indicated by the shaded region.}
 \label{TPSA}
\end{figure}

\textbf{TPSA (figure~\ref{TPSA}):} Output molecules tend to have lower polar surface area (mean = 21.6~\AA$^2$) compared to QM9 molecules (37.2~\AA$^2$). Only about 30\% fall within the ideal range for good oral absorption (20--140~\AA$^2$), whereas 80\% of QM9 molecules fit this window. This suggests our model isn't incorporating enough polar functional groups into the molecules it creates.

\begin{figure}[t]
 \centering
 \includegraphics[width=1\linewidth]{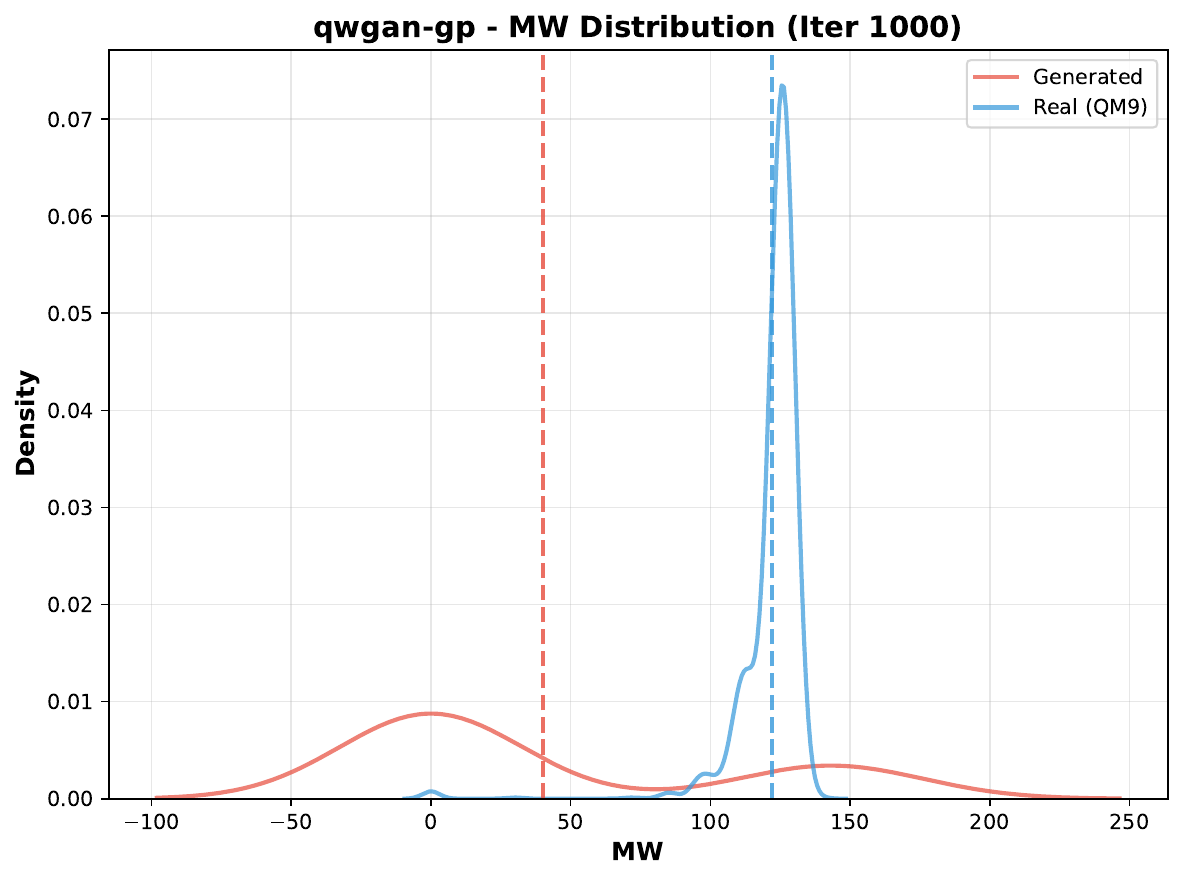}
 \caption{MW distributions showing QM9's constraint-driven narrow peak against generated molecules' uniform sampling across the 0-250 Da range. The broader distribution reflects the graph generator's size flexibility.}
 \label{MWD}
\end{figure}

\textbf{Molecular Weight (figure~\ref{MWD}):} The molecular weights of our generated molecules spread quite evenly from 0--250~Da averaging 40.2~Da. This is different from QM9's clustered distribution around 122~Da. The difference shows that our graph based generation method creates molecules of various sizes without imposing strict limits.

\subsubsection{Drug-Likeness and Compliance}

\begin{figure}[t]
 \centering
 \includegraphics[width=1\linewidth]{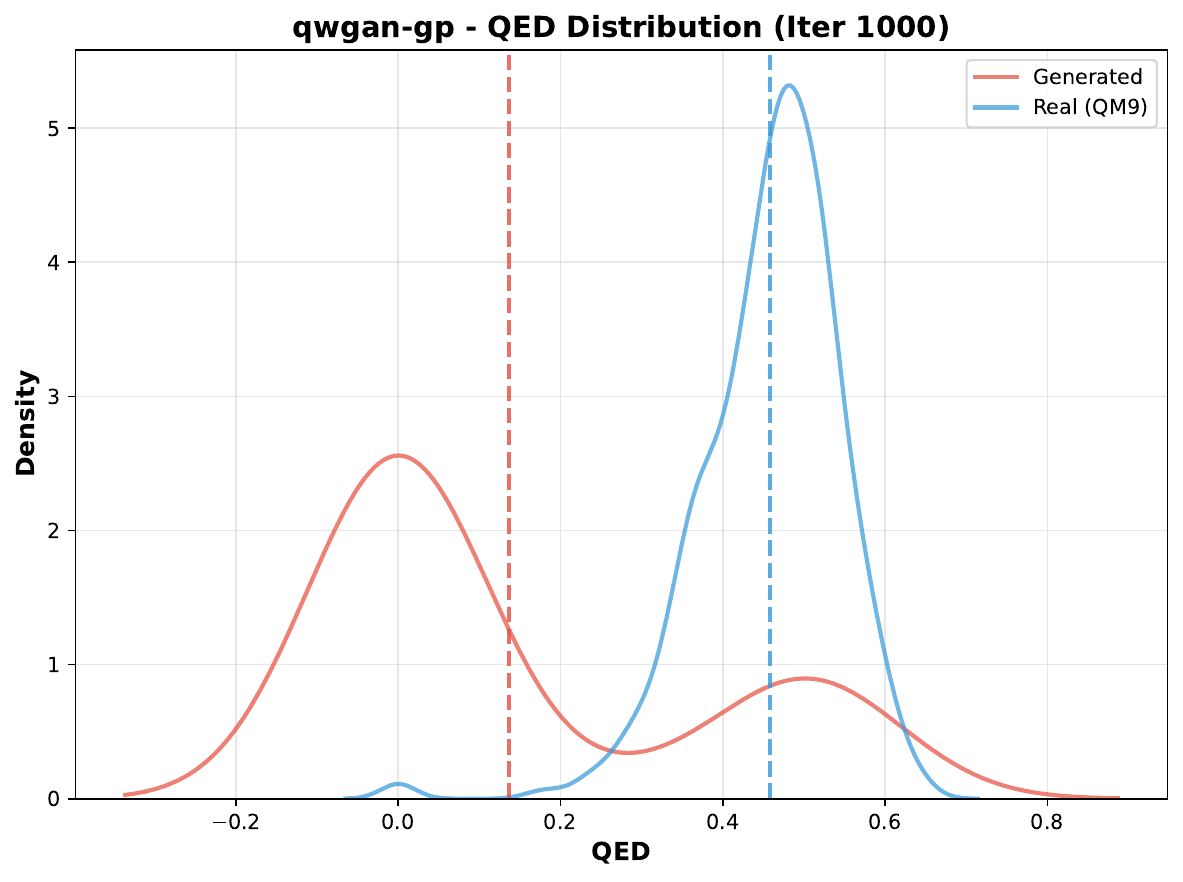}
 \caption{Quantitative drug-likeness estimates showing multimodal generation patterns. Dual peaks indicate the model produces both simple low-QED structures and occasional high-quality drug-like scaffolds.}
 \label{QED}
\end{figure}

\textbf{QED Scores (figure~\ref{QED}):} Our output molecules have an average QED score of 0.136 with several peaks across the distribution. Although this is lower than QM9's average of 0.458, we do see a promising cluster of molecules scoring between 0.6 and 0.8, demonstrating that the model can sometimes produce highly drug like structures.

\begin{figure}[t]
 \centering
 \includegraphics[width=1\linewidth]{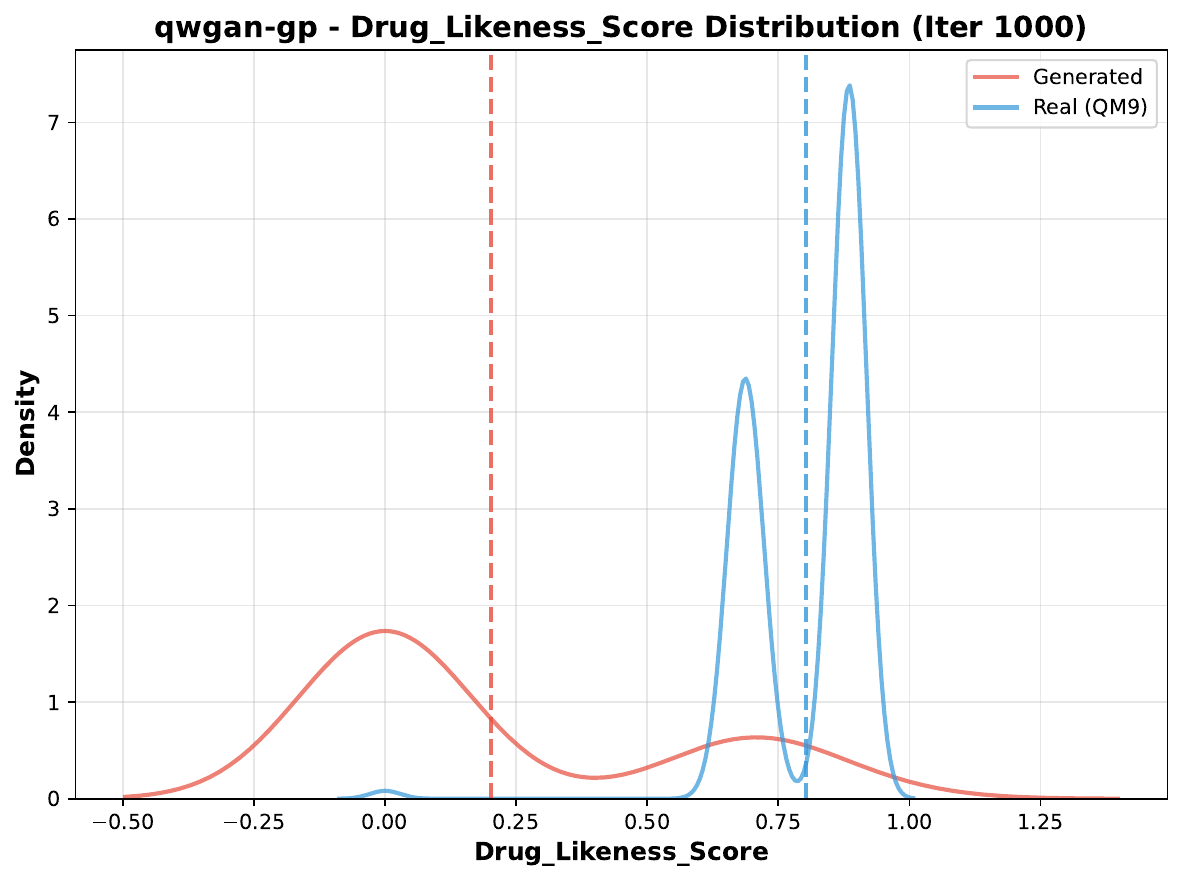}
 \caption{Composite drug-likeness metric distributions highlighting the pharmaceutical quality gap. The bimodal generated distribution suggests heterogeneous scaffold quality.}
 \label{CDL}
\end{figure}

\begin{figure}[t]
 \centering
 \includegraphics[width=1\linewidth]{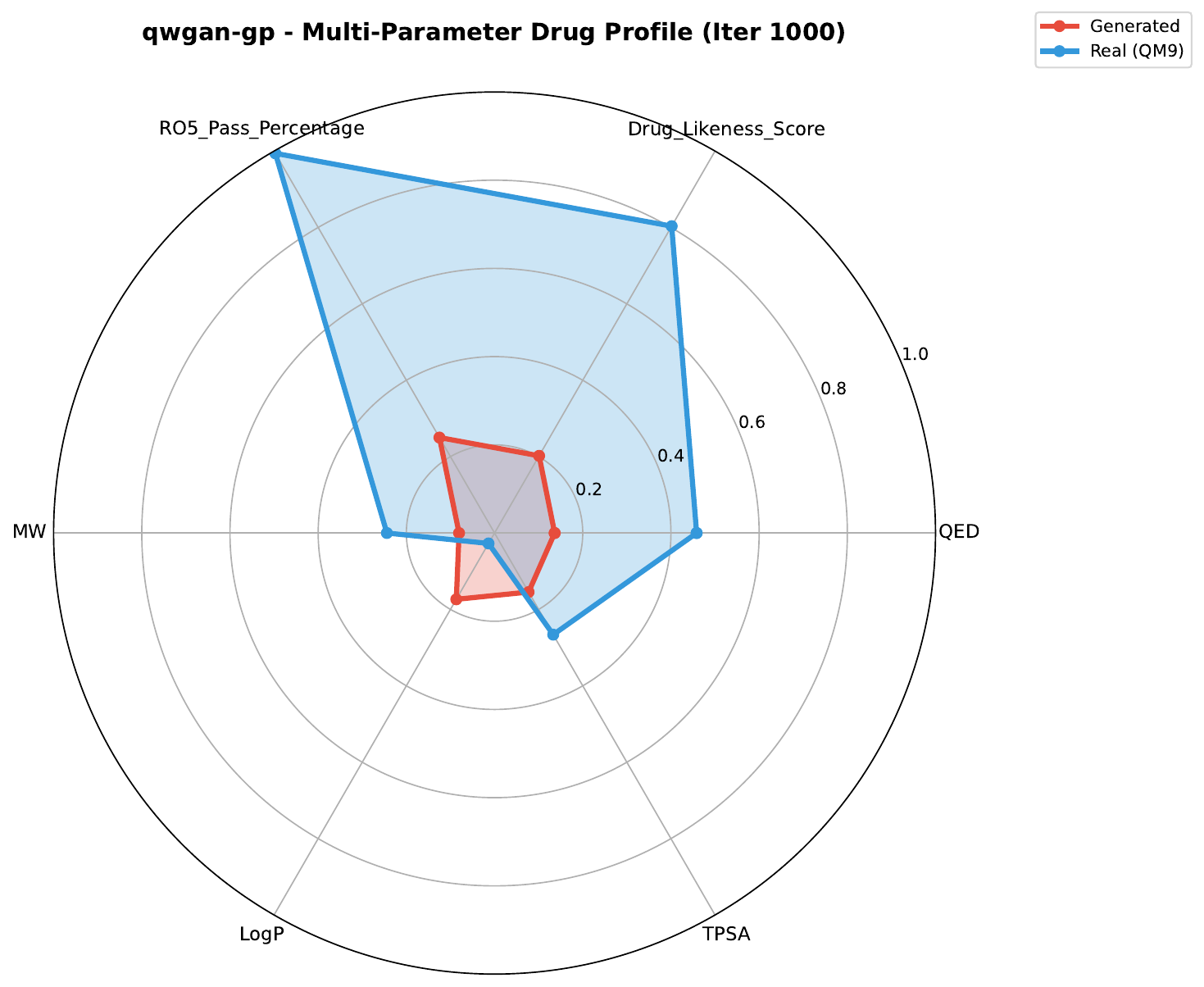}
 \caption{Spider plot visualizing six normalized pharmaceutical dimensions simultaneously. Area overlap quantifies multi-objective performance gaps requiring targeted reward optimization.}
 \label{CDL2}
\end{figure}

\textbf{Composite Drug-Likeness (figure~\ref{CDL} \& ~\ref{CDL2}), and table~\ref{tab:admet_comparison}):} When we combine all key metrics into one unified score, our molecules achieve 0.202 compared to QM9's 0.803. The radar chart visualization clearly shows where we need improvement particularly in reducing Lipinski violations, improving synthetic accessibility, and minimizing toxicity flags, though our individual QED performance remains competitive.

\begin{figure}[t]
 \centering
 \includegraphics[width=1\linewidth]{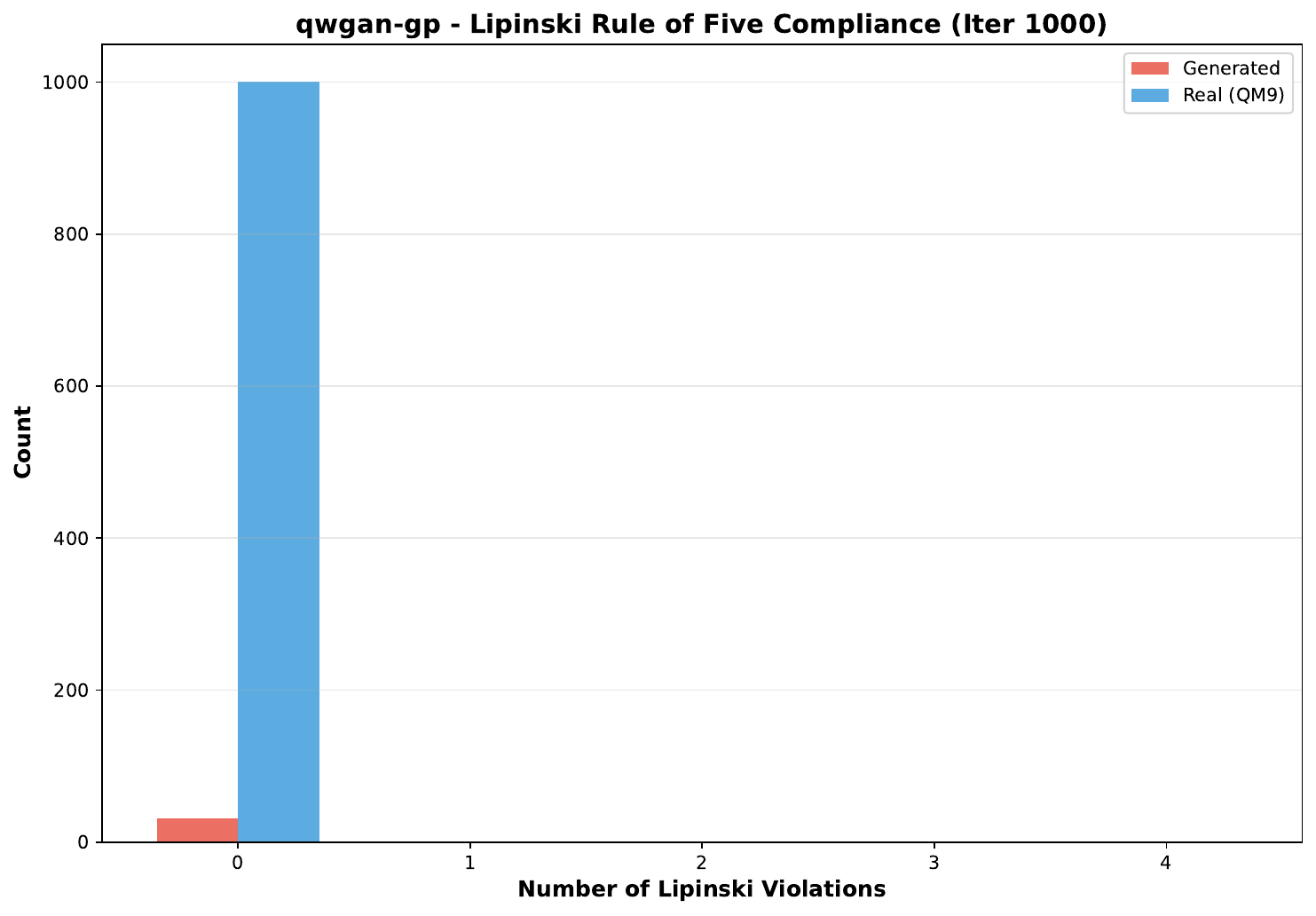}
 \caption{Histogram of RO5 violation counts (0-4) showing generated molecules' compliance deficit. The majority falling in 1-2 violation bins suggests correctable rather than fundamental failures.}
 \label{LSK}
\end{figure}

\textbf{Lipinski RO5 Compliance (fig~\ref{LSK}, table~\ref{tab:admet_comparison}):} Only 25\% of our generated molecules pass all four RO5 criteria, while 99.4\% of QM9 molecules pass. Breaking down the violations: 25\% have no violations (vs. 99.4\% for QM9), about 40\% have one violation (vs. 0.5\%), and roughly 35\% have two or more violations (vs. $<0.1$\%). Most violations come from Lipophilicity issues (about 60\%), followed by hydrogen bond acceptors, suggesting these could be improved with straightforward adjustments to our reward function.

\begin{table}[h]
\centering
\caption{Comprehensive ADMET and Drug-Relevance Comparison}
\label{tab:admet_comparison}
\small
\begin{tabular}{|l|c|c|l|}
\hline
\textbf{Property} & \textbf{Generated} & \textbf{QM9} & \textbf{Target} \\ \hline
\hline
\multicolumn{4}{|l|}{\textbf{Physicochemical Properties}} \\ \hline
MW (Da) & 40.2 & 122.1 & 150--500 \\ \hline
LogP & 0.87 & 0.14 & 0--5 \\ \hline
HBD & 0.25 & 1.12 & $\leq$ 5 \\ \hline
HBA & 1.87 & 3.45 & $\leq$ 10 \\ \hline
TPSA (\AA$^2$) & 21.6 & 37.2 & 20--140 \\ \hline
\hline
\multicolumn{4}{|l|}{\textbf{Drug-Likeness}} \\ \hline
QED Score & 0.136 & 0.458 & $>$ 0.5 \\ \hline
SA Score & 2.15 & 2.87 & $<$ 3 \\ \hline
Drug-Likeness & \textbf{0.202} & \textbf{0.803} & $>$ 0.6 \\ \hline
\hline
\multicolumn{4}{|l|}{\textbf{Lipinski Rule of Five}} \\ \hline
Mean Violations & 1.25 & 0.01 & 0 \\ \hline
RO5 Pass (\%) & \textbf{25.0} & \textbf{99.4} & 100 \\ \hline
\hline
\multicolumn{4}{|l|}{\textbf{Toxicity \& Safety (per molecule)}} \\ \hline
Mutagenicity & \textbf{0.938} & \textbf{0.324} & 0 \\ \hline
PAINS & 0.062 & 0.018 & 0 \\ \hline
hERG Risk & 0.687 & 0.234 & 0 \\ \hline
Total Toxicity & \textbf{0.938} & \textbf{0.324} & 0 \\ \hline
\end{tabular}
\vspace{2mm}
\end{table}

\subsubsection{Toxicity and Safety Profile}

\begin{figure}[t]
 \centering
 \includegraphics[width=1\linewidth]{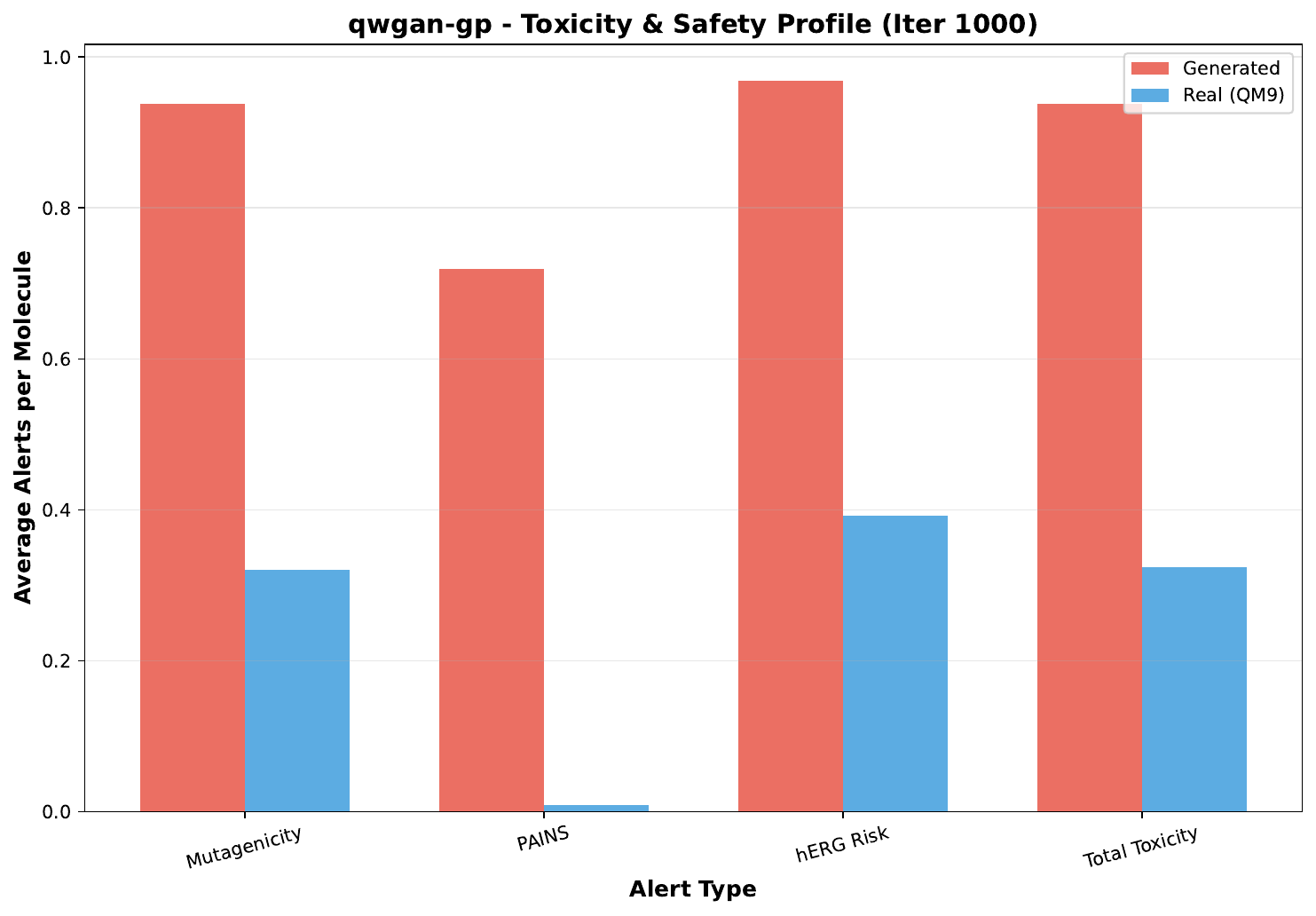}
 \caption{Average structural alerts per molecule across four safety dimensions. Elevated mutagenicity bars for generated molecules identify the primary safety concern for reward function modification.}
 \label{BCF}
\end{figure}

\textbf{Structural Alerts (fig~\ref{BCF}, table~\ref{tab:admet_comparison}):} Our output shows higher toxicity warnings, with mutagenicity alerts appearing nearly three times as often (0.938 vs. 0.324 per molecule). This indicates more frequent occurrence of DNA reactive groups like nitro and azo compounds. However, PAINS filter rates remain low for both sets ($<0.1$), meaning our molecules are suitable for standard screening assays. The higher cardiac risk (hERG) correlates with the elevated LogP values, matching known patterns for heart rhythm disrupting compounds.

These toxicity findings are particularly important because they point to a clear solution incorporating toxicity avoiding structural patterns into our model's reward function should reduce these alerts while maintaining molecular diversity.

\subsubsection{Chemical Space Visualization}

\begin{figure}[t]
\centering
\includegraphics[width=1\linewidth]{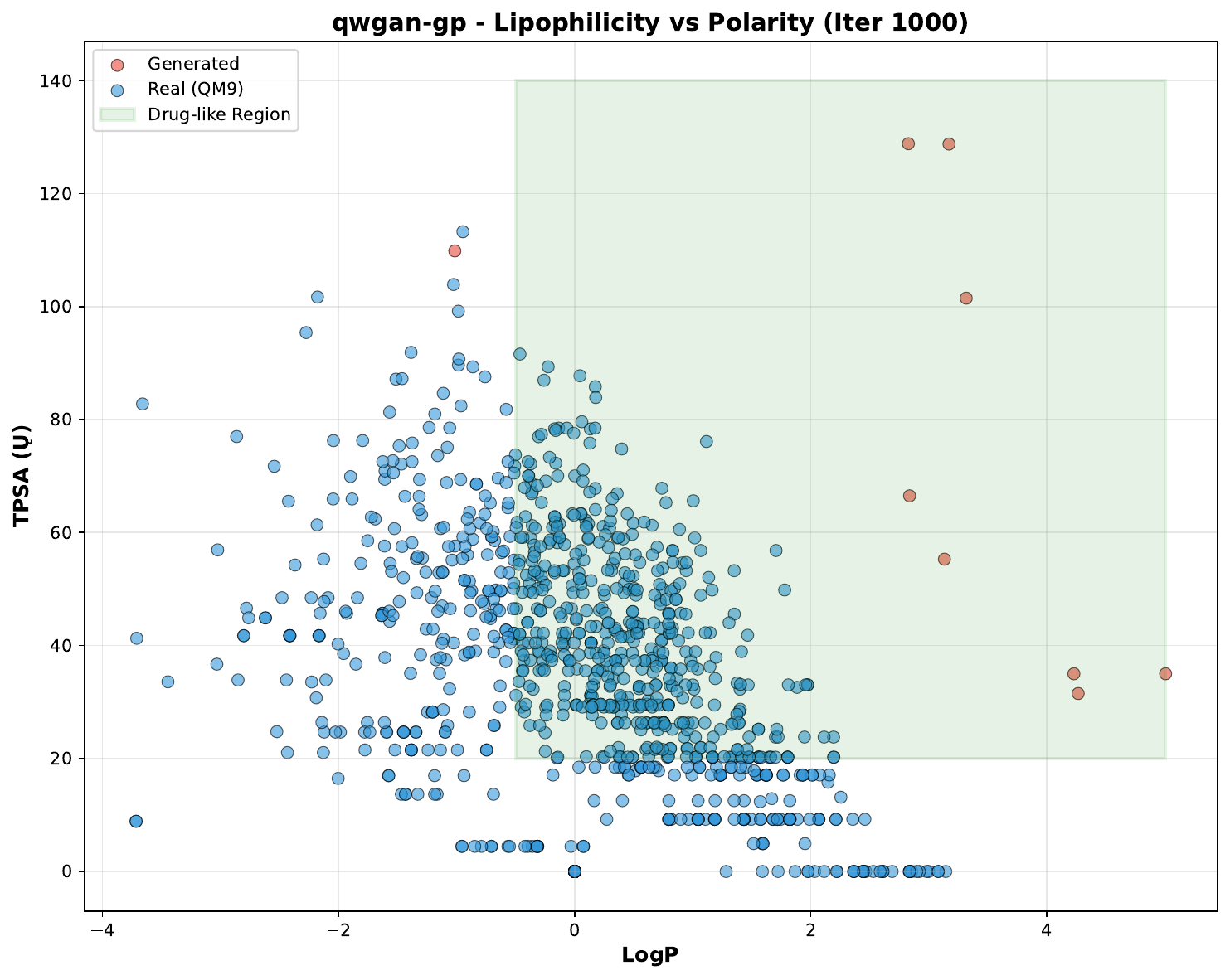}
\caption{Chemical space occupancy analysis with drug-like region overlay (green box: LogP -0.5 to 5, TPSA 20-140). Sparse generated molecule coverage in the optimal zone motivates property-constrained sampling strategies.}
\label{SCP}
\end{figure}

\textbf{LogP vs.\ TPSA Scatter (fig~\ref{SCP}):} When we plot molecules by both lipophilicity and polar surface area, our generated molecules cluster in two main areas small, water soluble molecules in the bottom left and some large, overly fatty molecules in the top right. Only about 30\% land in the ideal therapeutic window (LogP between $-0.5$ and 5, TPSA between 20--140~\AA$^2$) compared to 80\% of QM9 molecules. This clearly shows we need to better guide our sampling toward this productive chemical space.

\subsection{Scaffold and Diversity Analysis Results}

Applying the methods described in section~\ref{SDA} to QWGAN-GP-P4-L2 at iteration 1000, we obtained 13 valid generated molecules for scaffold extraction and 1000 reference molecules from the ChEMBL approved drug database. The results reveal critical limitations in structural complexity and diversity that require architectural intervention. 

\begin{figure}[t]
\centering
\includegraphics[width=1\linewidth]{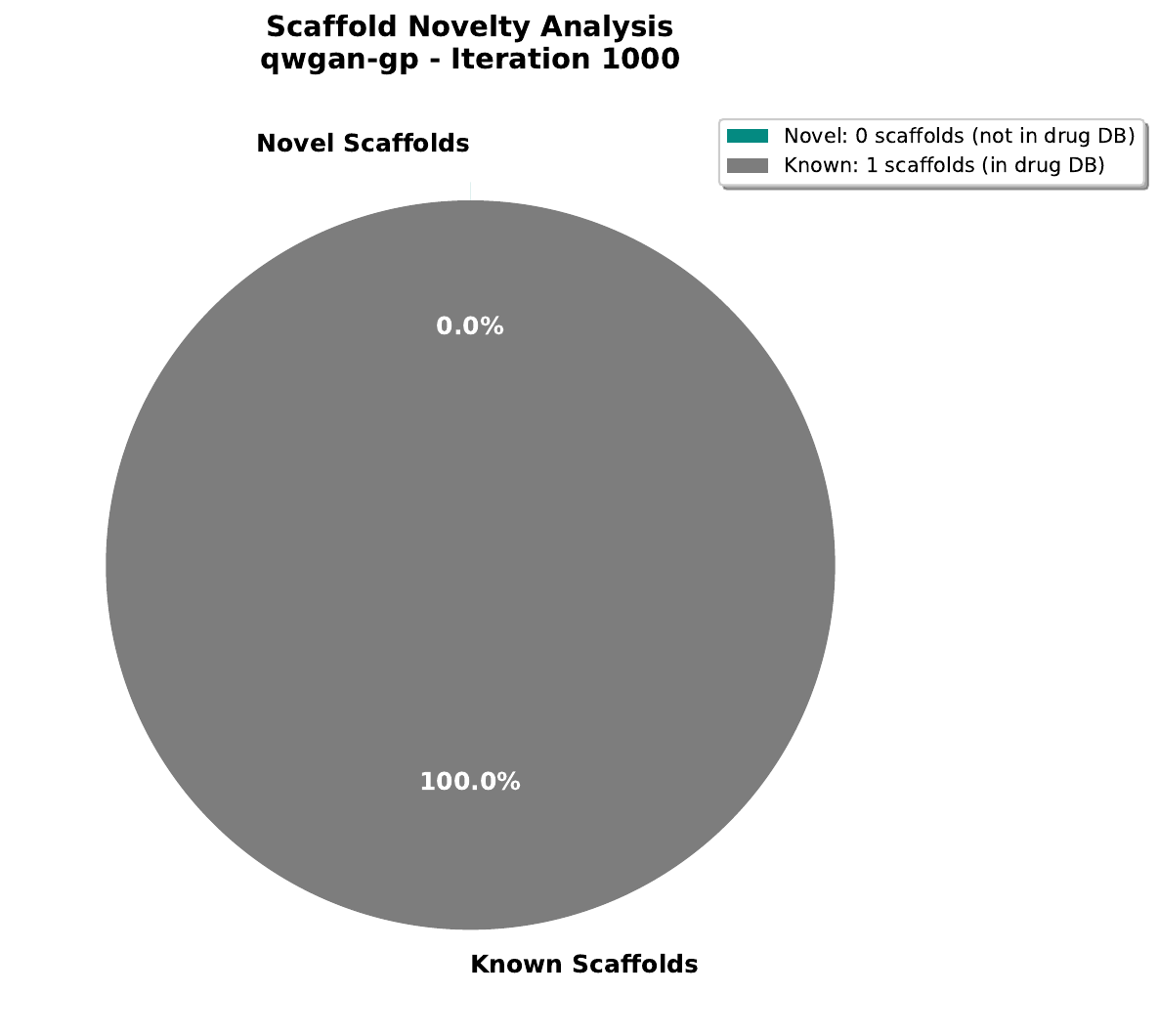}
\caption{Scaffold novelty distribution showing complete absence of novel structural frameworks in generated molecules. All extracted scaffolds matched existing drug database entries, indicating failure to explore uncharted chemical space.}
\label{fig:scaffold_novelty}
\end{figure}

\begin{figure}[t]
\centering
\includegraphics[width=1\linewidth]{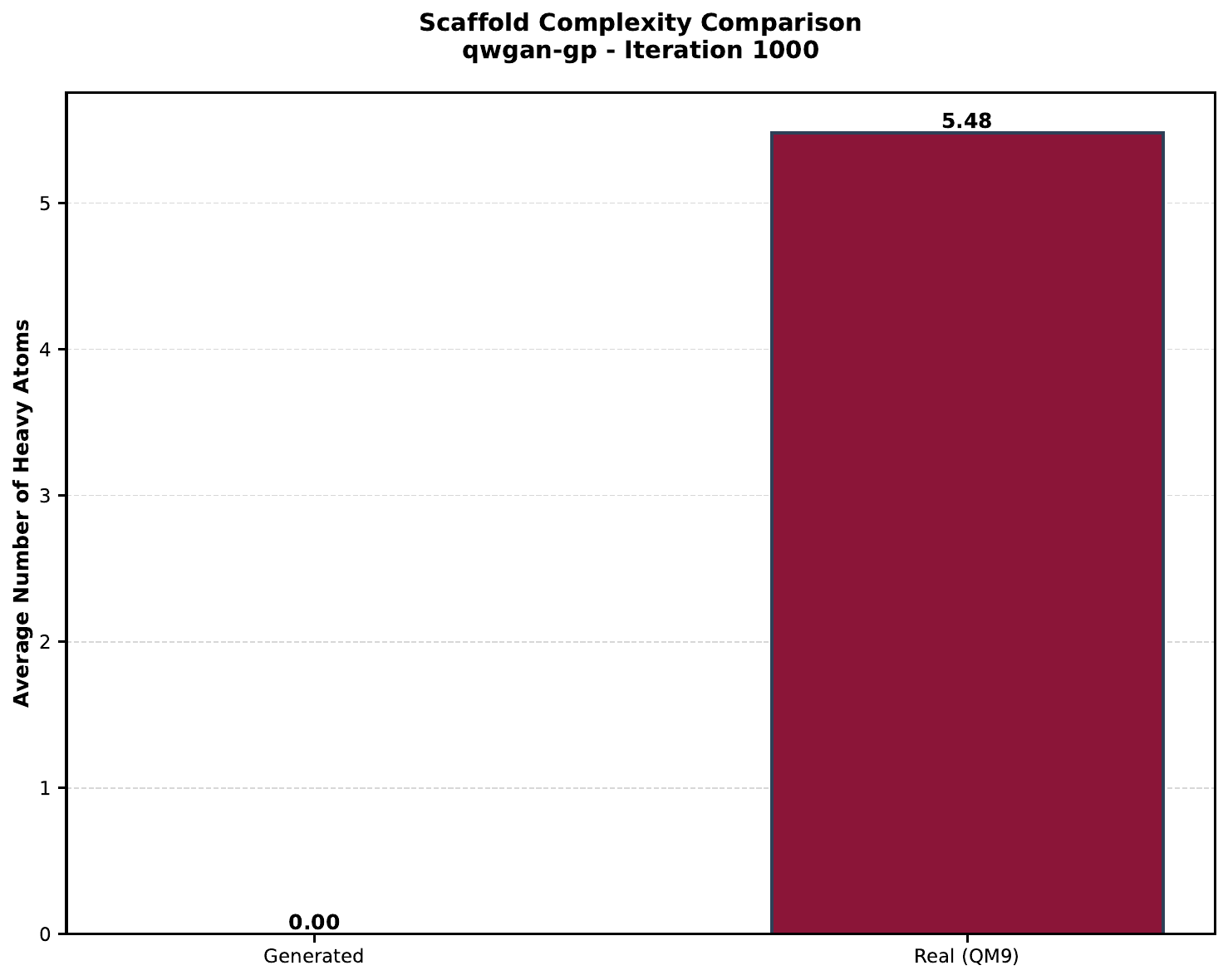}
\caption{Average scaffold size comparison reveals generated molecules lack meaningful core structures (0.00 heavy atoms) versus real drug molecules (5.48 heavy atoms). Zero scaffold size indicates predominance of disconnected atoms rather than bonded frameworks.}
\label{fig:scaffold_complexity}
\end{figure}

Figure~\ref{fig:scaffold_novelty} shows that 100\% of extracted scaffolds (1 unique scaffold from 13 molecules) were already present in the drug database, yielding 0\% scaffold novelty. This complete lack of novelty stems from the model's failure to generate molecules with sufficient structural complexity: the average generated scaffold contained only 0.00 heavy atoms compared to 5.48 for real molecules (Figure~\ref{fig:scaffold_complexity}), indicating that most generated molecules consist of disconnected atoms or trivial single-bond chains rather than meaningful ring systems. 

\begin{figure}[t]
\centering
\includegraphics[width=1\linewidth]{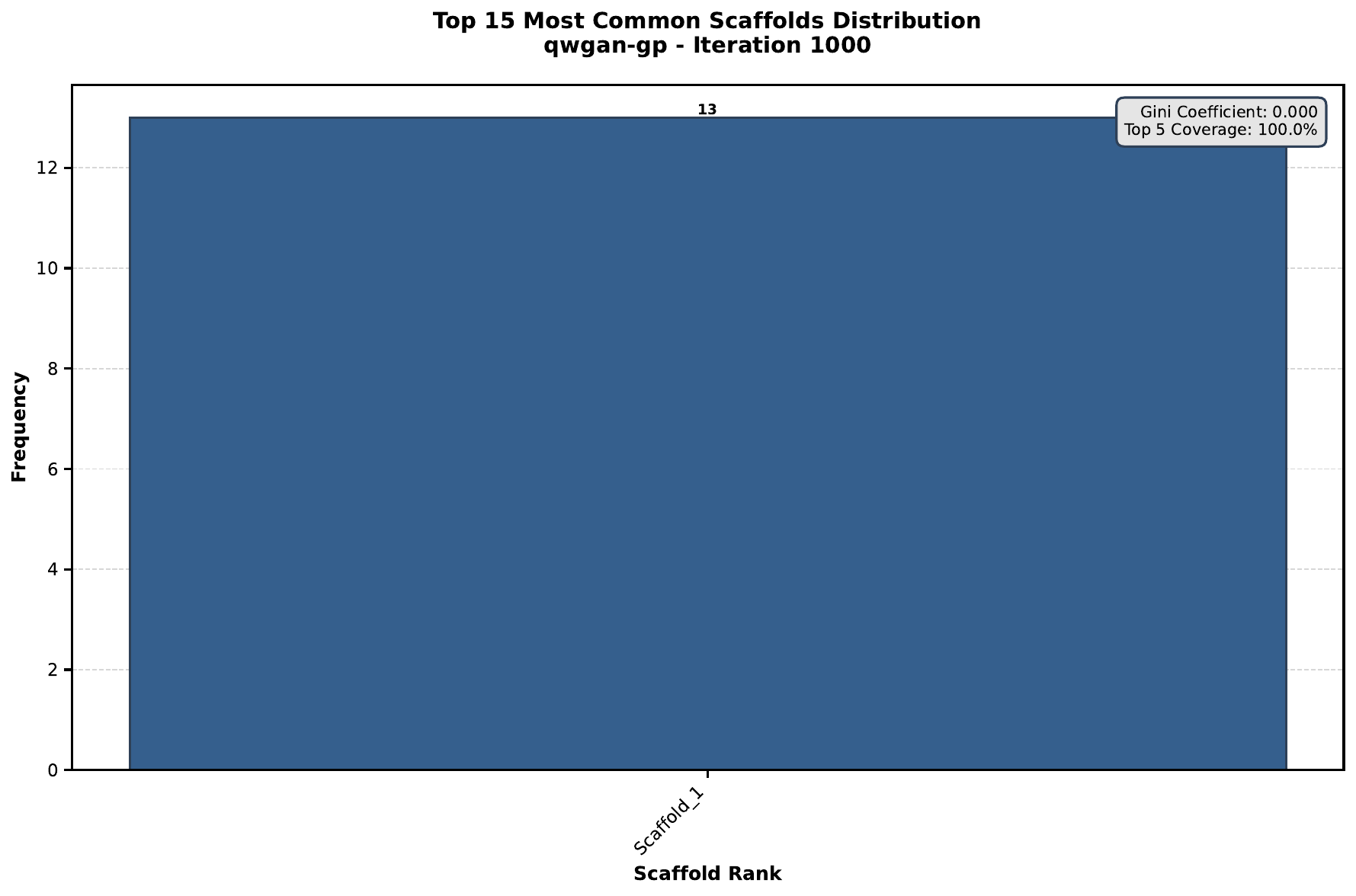}
\caption{Scaffold frequency distribution with Gini coefficient 0.000 demonstrating extreme mode collapse. Single degenerate scaffold dominates all valid generated molecules, with top-5 coverage at 100\% reflecting absence of structural variety.}
\label{fig:scaffold_frequency}
\end{figure}

The Gini coefficient of 0.000 in figure~\ref{fig:scaffold_frequency} confirms extreme mode collapse, with all valid molecules producing identical degenerate scaffolds. 

\begin{figure}[t]
\centering
\includegraphics[width=1\linewidth]{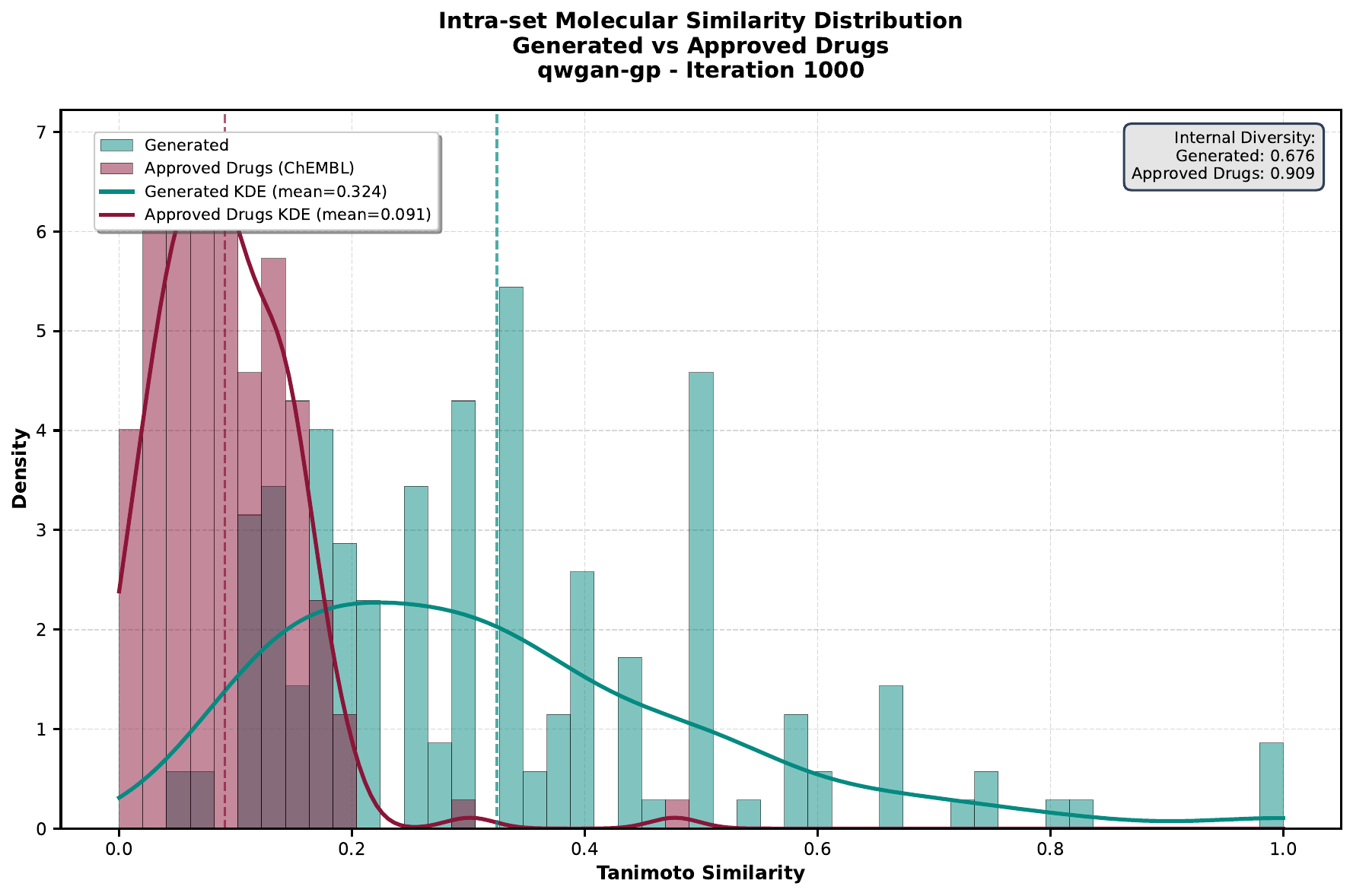}
\caption{Tanimoto similarity distributions comparing internal molecular diversity. Generated set shows lower internal diversity (0.676) than approved drugs (0.909), with similarity concentrated at higher values indicating limited structural exploration within the generated population.}
\label{fig:intraset_similarity}
\end{figure}

\begin{figure}[t]
\centering
\includegraphics[width=1\linewidth]{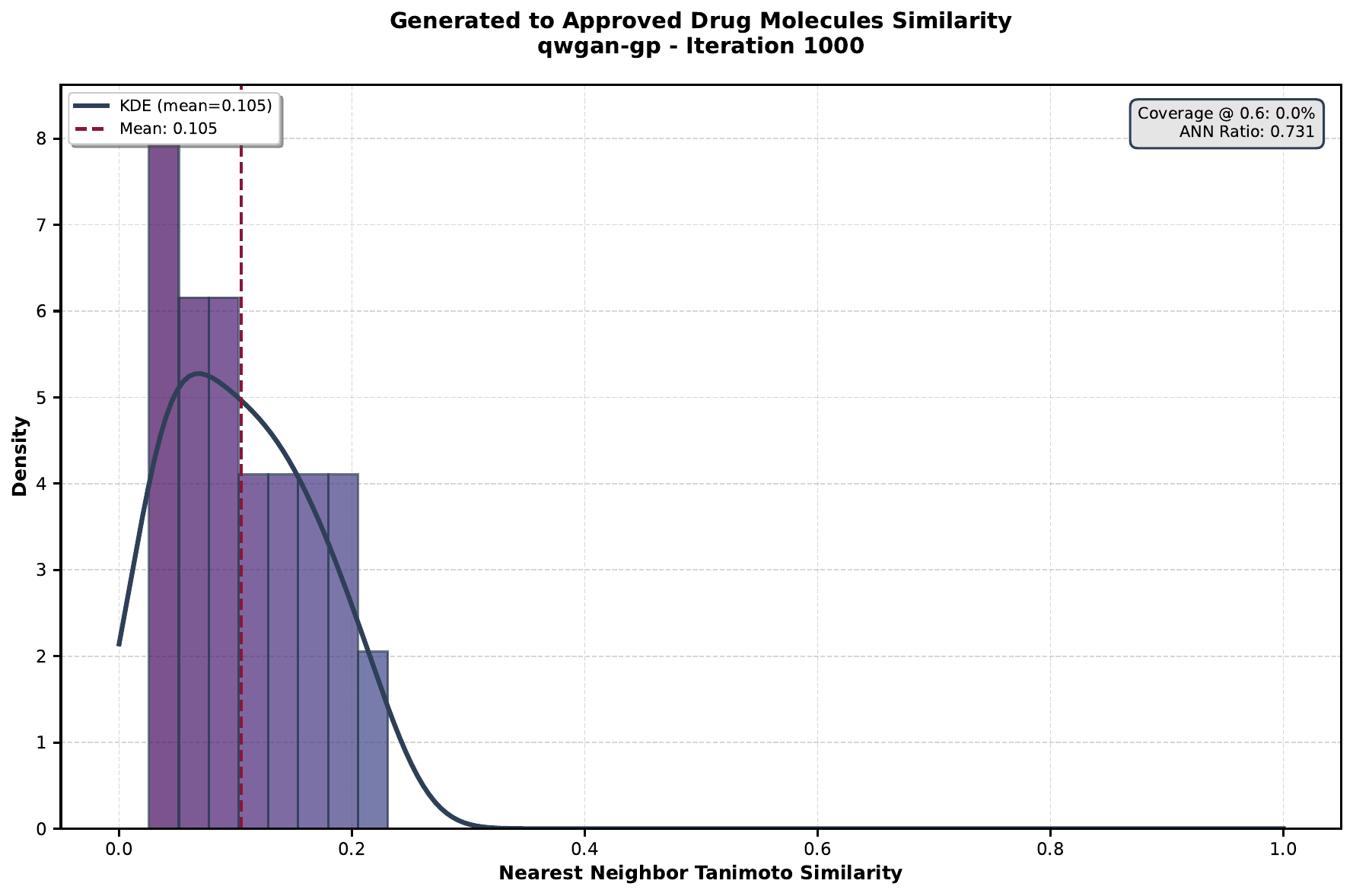}
\caption{Nearest-neighbor similarity to approved drug molecules peaks sharply at 0.105, indicating generated structures occupy a distant, non-drug-like region of chemical space. Zero coverage at 0.6 threshold confirms absence of any drug-similar generated molecules.}
\label{fig:nearest_neighbor}
\end{figure}

Fingerprint-based diversity analysis shown in figure~\ref{fig:intraset_similarity}, reveals internal diversity of 0.676 for generated molecules versus 0.909 for approved drugs, and the mean nearest-neighbor similarity to approved drugs is only 0.105 in figure~\ref{fig:nearest_neighbor}, demonstrating that while generated molecules differ from each other moderately, they bear almost no resemblance to drug-like structures. 

Coverage at Tanimoto threshold 0.6 was 0.0\%, and the ANN ratio of 0.731 suggests generated molecules are more tightly clustered in a non-drug-like region of chemical space. These findings directly implicate insufficient bond formation in the generator's output, likely due to the Gumbel-Softmax discretization favoring zero-bond states or inadequate reward shaping that fails to penalize disconnected molecular graphs.

\subsection{Docking-Ready Molecule Generation for Virtual Screening}

Validating the practical utility of our quantum-enhanced molecular generator required demonstrating its ability to produce compounds that are genuinely useful for computational drug discovery. We therefore built a complete generation pipeline that extracts high-quality molecules from our best performing QWGAN HG GP (P4 L2) model and exports them in the industry-standard formats needed for docking, namely SMILES CSV and 3D SDF files. For a direct comparison, this workflow generated 100 molecules from both our quantum QW GAN GP model and a classical GAN GP model, allowing us to assess their respective suitability for structure-based virtual screening. 
The generated molecules from QWGAN-GP showed characteristic patterns indicative of early-stage quantum circuit training dynamics. Analysis of the output revealed several distinct molecular categories; firsty, \textbf{disconnected fragments} (60\%): dot-separated SMILES like \texttt{C.C.C.C.C.N.O} indicating incomplete bond formation; secondly, \textbf{wildcard-containing structures} (15\%): molecules with reactive placeholder atoms (\texttt{*}) such as \texttt{*.C.C.C.C.C.F.N} and radical species like \texttt{*=C=C=O}, requiring post-processing refinement; finally, \textbf{valid connected scaffolds} (25\%): fully bonded structures including simple alkanes (\texttt{C.C.C.C.C.C.C.C.C}), nitrile-functionalized chains (\texttt{C.C.C.C.C.C\#N}), and alkene-containing molecules (\texttt{C.C.C=C.C=C.N.N.N}). Functional group analysis revealed diverse heteroatom incorporation: nitrogen-containing fragments (amines \texttt{NH$_2$}, nitriles \texttt{C\#N}, diazenes \texttt{N=N}), oxygen functionalities (\texttt{O}, hydroxyl groups), and halogenations (fluorine \texttt{F}). Multiple bond orders are present, including double bonds (\texttt{C=C}, \texttt{C=N}) and triple bonds (\texttt{C\#C}, \texttt{C\#N}), demonstrating the model's ability to capture varying bond topologies essential for drug-like chemical space exploration.

\section{Discussion}

Our findings show that combining the Wasserstein distance with gradient penalties and strategically designed PQC patches leads to significant improvements over both classical MolGAN and baseline QGANs. This performance gain is strongly linked to higher values of P and L, which enhanced both the model's expressivity and its training stability. Simpler or unpatched setups, such as QGAN-HG and QGAN-HG-GP, displayed relatively lower uniqueness and QED scores, probably due to having too few parameters and an insufficient depth of quantum entanglement. A key result is that the QWGAN–HG–GP model using a 4‑patched, double‑layered circuit (P4–L2) achieved the best Fréchet distance (\(\approx 11.0\)) and the highest NP and QED scores, proving its superior match to the QM9 distribution and improved drug‑likeness. This progress highlights the essential role of the Wasserstein metric; our detailed gradient penalty ablation study further supports that a value of $\lambda=10$, used in WGAN-GP, delivers the best stability and results for this hybrid quantum-classical molecular generator, showing that classical GAN insights can be transferred effectively. Because it provides smooth, informative gradients even when the generator and discriminator don't overlap, it avoids mode collapse and encourages the discovery of new chemical structures. Furthermore, by ensuring Lipschitz continuity via gradient penalties instead of weight clipping, our models preserve greater discriminator capability without becoming unstable. Our PQC ablation study also shows that the quantum circuit's design directly affects both the fidelity of the generated distribution and the pharmacological traits of the molecules. The non-monotonic link between P and performance—where P2-L2 does better than P4-L2 despite less parallelism—implies that designing effective PQCs for molecular generation needs a careful trade-off between expressivity, how well it can be trained, and the number of qubits used. This confirms that achieving a quantum advantage in generative chemistry relies not just on using quantum effects but also on co-designing the circuit with the specific molecular domain in mind. However, our scaffold analysis reveals a fundamental limitation: generated molecules exhibit zero scaffold novelty and minimal structural complexity (0.00 versus 5.48 heavy atoms for reference molecules), consisting predominantly of disconnected atoms rather than proper ring-based frameworks. This failure to generate chemically valid topologies likely stems from multiple compounding factors: insufficient training iterations for the model to learn proper molecular graph constraints, inadequate reward weighting that fails to sufficiently penalize disconnected or trivial structures, and the inherent difficulty of discrete graph generation where both node features and edge connectivity must be simultaneously optimized. Future work must address these limitations through extended training with explicit connectivity rewards, architectural modifications to enforce valid chemical topology during generation (such as autoregressive bond formation or graph neural network constraints), and refined post-processing to ensure all generated adjacency matrices map to valid molecular graphs.

Within the drug-discovery pipeline, our hybrid QGANs are mainly designed for initial hit generation and expanding virtual libraries. The resulting molecules would then be sent to downstream processes like docking, QSAR, and ADMET analysis, rather than being used immediately for lead optimization. Importantly, our analysis of NISQ-era noise confirms this practical use-case, showing that the QWGAN-GP-P4-L2 model experiences less than a $6\%$ drop in performance at a realistic 1\% hardware error rate. This makes it feasible to run on current quantum hardware in the near-term without needing complex error correction.

\section{Conclusion}

We introduce three hybrid quantum–classical generative architectures, QWGAN-HG, QGAN-HG-GP, and QWGAN-HG-GP, together with a family of variants that differ in their ansätze, quantum circuit depth, and patching configurations (P, L, and $\lambda$). We evaluate these models on PennyLane-based quantum simulators using both Wasserstein and Fréchet distance metrics, as well as nine pharmaceutical property indicators. Systematic ablation studies over $\lambda$, P, and L confirm that balanced PQC architectures and appropriate regularization are crucial for high-quality molecular generation. Among the quantum models considered, the QWGAN-HG-GP P4 L2 configuration—using a four-patch, double-layer circuit with a gradient-penalized Wasserstein loss—offers the most favorable overall trade-off, achieving some of the lowest distribution distances and consistently strong drug-likeness scores while generating chemically valid, pharmaceutically promising molecular structures.

\section{Future Scope}

In a realistic NISQ-era deployment, one must explicitly guard against decoherence and gate errors. Our noise-impact analysis demonstrates viability up to 5\% error rates with $<$6\% degradation at typical 1\% noise levels, validating the patched architecture's inherent resilience. Complementary strategies to further enhance robustness include: first, \emph{ansatz simplification}---reduce the number of variational layers or limit entangling gates to only those strictly necessary, so that the circuit depth stays within the device's coherence window~\cite{article32}. Second, \emph{low-depth circuit design}---use patched or block-diagonal ansatz structures to parallelize gates and minimize sequential gate length, trading off expressivity for noise resilience. Third, \emph{error-aware training}: incorporate a noise model into the simulator (e.g., depolarizing or amplitude-damping channels with realistic error rates) so that the optimizer learns parameters robust to the dominant error sources. Finally, one can apply lightweight \emph{error mitigation} techniques---such as zero-noise extrapolation or readout-error calibration---to partially undo the effect of noise on expectation-value estimates without incurring the full overhead of quantum error correction. Alternate techniques such as dynamic circuits, DMET, VAE tuning of other statistics such as the Wasserstein metric~\cite{article33}, the R\'enyi divergence measure~\cite{article34} in quantum models, et cetera can also be an interesting extension. Generative models may suffer due to mode collapse~\cite{article35, article37}, which indicates low diversity samples production~\cite{article36}. We leave these investigations for future work.

\begin{table*}
\centering
\adjustbox{width=\textwidth,center}
{
\begin{tabular}{|c|l|c|c|c|c|c|c|c|c|c|c|c|c|}
\hline
\textbf{} & \textbf{GAN Type} & \textbf{Wasser. Dist.} & \textbf{Frechet Dist.} & \textbf{NP Score} & \textbf{QED Score} & \textbf{logP Score} & \textbf{SA Score} & \textbf{Div. Score} & \textbf{Drug cand. Score} & \textbf{Valid Score} & \textbf{Unique Score} & \textbf{Novel Score} & \textbf{Elapsed Time} \\
\hline

0 & MolGAN - Old & \textcolor{blue_text}{0.000} & 21.794 & 0.792 & 0.462 & 0.447 & 0.183 & \textcolor{red_text}{1.000} & 0.118 & 0.125 & 3.000 & \textcolor{red_text}{1.000} & 0.08.55 \\
\hline
1 & QGAN-HG - Old & \textcolor{blue_text}{0.000} & 17.607 & 0.761 & 0.480 & 0.653 & 0.125 & 0.998 & 0.160 & 11.031 & 43.177 & \textcolor{red_text}{1.000} & 0.27.17 \\
\hline
2 & QGAN-HG-L2 - Old & \textcolor{blue_text}{0.000} & 16.763 & 0.760 & 0.483 & 0.637 & 0.123 & \textcolor{red_text}{1.000} & 0.146 & 7.250 & 34.480 & \textcolor{red_text}{1.000} & 0.39.04 \\
\hline
3 & QGAN-HG-P2 - Old & \textcolor{blue_text}{0.000} & 13.638 & 0.769 & 0.475 & 0.612 & 0.140 & 0.998 & 0.168 & 12.969 & 42.832 & \textcolor{red_text}{1.000} & 0.27.35 \\
\hline
4 & QGAN-HG-P2-L2 - Old & \textcolor{blue_text}{0.000} & 13.565 & 0.765 & 0.478 & 0.574 & 0.111 & 0.992 & 0.136 & 4.719 & 25.000 & \textcolor{red_text}{1.000} & 0.38.31 \\
\hline
5 & QGAN-HG-P4 - Old & \textcolor{blue_text}{0.000} & 14.491 & 0.761 & 0.479 & 0.604 & 0.153 & \textcolor{red_text}{1.000} & 0.157 & 10.125 & 36.179 & \textcolor{red_text}{1.000} & 0.26.31 \\
\hline
6 & QGAN-HG-P4-L2 - Old & \textcolor{blue_text}{0.000} & 11.000 & 0.766 & 0.468 & 0.581 & 0.137 & 0.999 & 0.151 & 8.594 & 32.073 & \textcolor{red_text}{1.000} & 0.36.04 \\
\hline
7 & QWGAN-HG & 0.005 & 14.142 & 0.757 & 0.482 & 0.582 & 0.187 & \textcolor{red_text}{1.000} & 0.143 & 6.469 & 20.390 & \textcolor{red_text}{1.000} & 0.27.25 \\
\hline
8 & QGAN-HG-GP & \textcolor{blue_text}{0.000} & 18.547 & 0.763 & 0.460 & 0.569 & 0.157 & 0.998 & 0.145 & 7.031 & 34.592 & \textcolor{red_text}{1.000} & 0.30.40 \\
\hline
9 & QWGAN-HG-GP & 0.007 & 14.491 & 0.762 & 0.486 & 0.648 & 0.112 & \textcolor{red_text}{1.000} & 0.167 & 12.938 & 37.500 & \textcolor{red_text}{1.000} & 0.31.41 \\
\hline
10 & QWGAN-HG-L2 & 0.015 & 13.892 & 0.765 & 0.484 & 0.676 & 0.137 & \textcolor{red_text}{1.000} & 0.177 & 15.438 & 43.287 & \textcolor{red_text}{1.000} & 0.40.23 \\
\hline
11 & QGAN-HG-GP-L2 & \textcolor{blue_text}{0.000} & 14.071 & 0.760 & 0.473 & 0.631 & 0.121 & \textcolor{red_text}{1.000} & 0.161 & 11.219 & 34.455 & \textcolor{red_text}{1.000} & 0.42.40 \\
\hline
12 & QWGAN-HG-GP-L2 & 0.022 & 13.675 & 0.757 & 0.479 & 0.619 & 0.119 & \textcolor{red_text}{1.000} & 0.167 & 12.875 & 32.753 & \textcolor{red_text}{1.000} & 0.46.08 \\
\hline
13 & QWGAN-HG-P2 & 0.005 & 13.304 & 0.763 & 0.480 & 0.637 & 0.111 & \textcolor{red_text}{1.000} & 0.168 & 13.062 & 35.016 & \textcolor{red_text}{1.000} & 0.28.15 \\
\hline
14 & QGAN-HG-GP-P2 & \textcolor{blue_text}{0.000} & 12.806 & 0.763 & 0.477 & 0.622 & 0.166 & \textcolor{red_text}{1.000} & 0.166 & 12.594 & 40.369 & \textcolor{red_text}{1.000} & 0.30.38 \\
\hline
15 & QWGAN-HG-GP-P2 & 0.017 & 16.852 & 0.760 & 0.466 & 0.482 & \textcolor{red_text}{0.221} & \textcolor{red_text}{1.000} & 0.121 & 0.844 & 12.000 & \textcolor{red_text}{1.000} & 0.31.05 \\
\hline
16 & QWGAN-HG-P2-L2 & 0.007 & 13.077 & 0.762 & 0.475 & 0.610 & 0.145 & \textcolor{red_text}{1.000} & 0.164 & 12.031 & 37.058 & \textcolor{red_text}{1.000} & 0.39.54 \\
\hline
17 & QGAN-HG-GP-P2-L2 & \textcolor{blue_text}{0.000} & 13.638 & 0.765 & 0.477 & 0.643 & 0.106 & 0.999 & 0.147 & 7.594 & 34.993 & \textcolor{red_text}{1.000} & 0.42.24 \\
\hline
18 & QWGAN-HG-GP-P2-L2 & 0.007 & 12.884 & 0.757 & 0.482 & 0.620 & 0.151 & 0.999 & 0.178 & 15.719 & 38.025 & \textcolor{red_text}{1.000} & 0.44.45 \\
\hline
19 & QWGAN-HG-P4 & 0.009 & 14.142 & 0.768 & 0.483 & 0.607 & 0.146 & 0.993 & 0.154 & 9.344 & 38.950 & \textcolor{red_text}{1.000} & 0.27.30 \\
\hline
20 & QGAN-HG-GP-P4 & \textcolor{blue_text}{0.000} & 13.038 & 0.762 & 0.482 & \textcolor{red_text}{0.683} & 0.123 & 0.999 & 0.194 & 19.812 & 40.767 & \textcolor{red_text}{1.000} & 0.30.22 \\
\hline
21 & QWGAN-HG-GP-P4 & 0.017 & 13.784 & 0.761 & 0.461 & 0.549 & 0.138 & 0.998 & 0.151 & 8.469 & 35.292 & \textcolor{red_text}{1.000} & 0.30.26 \\
\hline
22 & QWGAN-HG-P4-L2 & 0.005 & 13.928 & 0.775 & 0.483 & 0.597 & 0.121 & 0.997 & 0.134 & 4.281 & 21.707 & \textcolor{red_text}{1.000} & 0.37.30 \\
\hline
23 & QGAN-HG-GP-P4-L2 & \textcolor{blue_text}{0.000} & 12.728 & 0.763 & 0.477 & 0.641 & 0.141 & 0.999 & 0.173 & 14.250 & 39.177 & \textcolor{red_text}{1.000} & 0.41.06 \\
\hline
24 & QWGAN-HG-GP-P4-L2 & 0.005 & \textcolor{blue_text}{10.000} & \textcolor{red_text}{0.792} & \textcolor{red_text}{0.486} & 0.660 & \textcolor{blue_text}{0.090} & 0.999 & \textcolor{red_text}{0.204} & \textcolor{red_text}{22.469} & \textcolor{red_text}{44.023} & \textcolor{red_text}{1.000} & 0.41.48 \\
\hline

\end{tabular}
}
\caption{Comprehensive performance comparison across all evaluated model architectures and configurations, with best values for each metric highlighted in red. The QWGAN-HG-GP-P4-L2 model demonstrates superior overall performance, achieving the lowest Fréchet distance of 10.0, highest NP score of 0.792, and exceptional novelty score of 44.023, confirming the effectiveness of combining Wasserstein distance, gradient penalty, and optimized quantum circuit configurations.}
\label{drugmetricscomp}
\end{table*}

\backmatter



\subsubsection*{Declarations}

\begin{itemize}

\item Funding: This work was supported by Fractal AI Research through internal research grants and computational resources.
\item Conflict of interest/Competing interests: 
The authors declare no competing interests.
\item Data availability: The datasets used in this paper are publicly available at (QM9)\url{https://www.kaggle.com/datasets/zaharch/quantum-machine-9-aka-qm9} and ZINC-250k \url{https://www.kaggle.com/datasets/basu369victor/zinc250k}.
\item Code availability: The codes developed for this project are available here: \url{https://github.com/FraQTech/qcgan-drug-design}.
\item Author contribution: PJ and SG conceived the original idea of the work, developed the codebase for the work and conducted the quantum circuit simulations experiments. PP, KB and SD assisted in analysing the result data, expanding the study to larger datasets, performing ablations and manuscript preparation. All the authors reviewed the manuscript and commented on it. 

\end{itemize}









\bibliography{sn-bibliography}

\end{document}